\pdfoutput=1

\documentclass[12pt]{article}
\usepackage{jheppub}

\usepackage{amsmath}
\usepackage{amssymb}
\usepackage{graphicx,color,slashed,braket}
\usepackage{bbm}
\usepackage{ifpdf}
\usepackage{comment}
\usepackage{float}
\usepackage{enumitem}
\usepackage{stmaryrd}

\newcommand{\bC}{\mathbb{C}}

  \newcommand{\cN}{\mathcal{N}}
 \newcommand{\cR}{\mathcal{R}} \newcommand{\cW}{\mathcal{W}}

\newcommand{\be}{\begin{equation}} \newcommand{\ee}{\end{equation}}
\newcommand{\ba}{\begin{eqnarray}} \newcommand{\ea}{\end{eqnarray}}
\newcommand{\lp}{\left(}
\newcommand{\rp}{\right)}

\newcommand{\w}{\wedge}

 \newcommand{\bZ}{\mathbb{Z}}

\def\tad{N_{\rm flux}}

\def\rmi{{\rm i}}   \def\ii{\rmi} \def\xx{{\mathbf{x}}}
\def\nn{{\mathbf{n}}} \def\kk{{\mathbf{k}}} \def\00{{\mathbf{0}}} \def\11{{\mathbf{1}}}
\def\22{{\mathbf{2}}} \def\33{{\mathbf{3}}} \def\ll{{\mathbf{l}}} \def\LL{{\mathbf{L}}}

\def\Re{\mathop{\rm Re}\nolimits} \def\Im{\mathop{\rm Im}\nolimits}

\makeatletter
\newcommand{\xleftrightarrow}[2][]{%
\ext@arrow 0055{\leftrightarrowfill@}{#1}{#2}%
} \def\leftrightarrowfill@{%
\arrowfill@\leftarrow\relbar\rightarrow%
}
\makeatother

\begin{document}

\title{Fully stabilized Minkowski vacua in the $2^6$ Landau-Ginzburg model}

\author[b]{Muthusamy Rajaguru,}
\emailAdd{muthusamy.rajaguru@lehigh.edu}
\author[a]{Anindya Sengupta,}
\emailAdd{anindya.sengupta@tamu.edu} 
\author[b]{and Timm Wrase}
\emailAdd{timm.wrase@lehigh.edu}

\affiliation[a]{George P.\ and Cynthia Woods Mitchell Institute for Fundamental Physics and
Astronomy \\
Texas A\&M University, College Station, TX 77843, USA}
\affiliation[b]{Department of Physics, Lehigh University, 16 Memorial Drive East, Bethlehem, PA 18018, USA}

\allowdisplaybreaks

\abstract{We study moduli stabilization via fluxes in the $2^6$ Landau-Ginzburg model. Fluxes not only give masses to scalar fields but can also induce higher order couplings that stabilize massless fields. We investigate this for several different flux choices in the $2^6$ model and find two examples that are violating the Refined Tadpole Conjecture by more than a factor of 2. We also present, to our knowledge, the first 4d $\mathcal{N}=1$ Minkowski solution in string theory without any flat direction.}

\noindent
\maketitle
\newpage

\section{Introduction}\label{sec:introduction} 
Moduli stabilization is arguably one of the most important challenges in string phenomenology. After tremendous progress since the early 2000's, many constructions and scenarios have been called into question in the last few years within the swampland program. This is not only true for dS vacua with a positive cosmological constant, but also for supersymmetric AdS vacua that give rise to low-energy effective theories whose AdS scale is separated from the KK-scale. Four-dimensional Minkowski vacua either contain massless scalar fields for $\mathcal{N}\geq 2$ or are expected to receive corrections that potentially invalidate the very existence of a solution for $\mathcal{N} \leq 1$. So, there seem to be currently no agreed-upon compactifications to lower dimensions without a moduli space.

Probably the simplest way of generating a scalar potential for moduli is by turning on fluxes in the internal space. Such fluxes in type II string theory induce generically a charge that needs to be canceled by O-planes. These O-planes together with the fluxes lead to a scalar potential that depends on all the K\"ahler moduli for Calabi-Yau (CY) compactifications of massive type IIA with smeared O6-planes \cite{DeWolfe:2005uu}. Similarly, it depends on all complex structure moduli for CY compactifications of type IIB in the presence of an O3/O7 orientifold \cite{Giddings:2001yu}. Rigid CY-manifolds have $h^{2,1}=0$ and therefore no complex structure moduli. They, therefore, allow for the stabilization of all moduli in massive type IIA but it has not yet been possible to find solutions with fully localized O6-planes, see \cite{Junghans:2020acz, Marchesano:2020qvg, Cribiori:2021djm, Emelin:2022cac, Junghans:2023yue, Bardzell:2024anh, Emelin:2024vug} for recent efforts in that direction. For type IIB on the other hand it is not too difficult to find fully localized solutions in which the backreaction of the O3-planes is encoded in a non-trivial warp factor \cite{Giddings:2001yu}. However, in that case, the stabilization of the K\"ahler moduli would have to rely on non-perturbative effects \cite{Kachru:2003aw}. 

An intriguing way to avoid the complication of having to deal with a combination of fluxes and non-perturbative effects was proposed in \cite{Becker:2006ks}. The mirror dual of a rigid CY-manifold has $h^{1,1}=0$ and therefore no K\"ahler moduli at all. That means one can stabilize \emph{all} moduli in type IIB by using fluxes only. While there is no geometric interpretation of the internal space in this case, one can study such setups using the language of Landau-Ginzburg (LG) models \cite{Vafa:1988uu,vafaOrbifoldized, Schimmrigk:1992ai, Witten:1993yc, Candelas:1993nd}. This was done almost twenty years ago in \cite{Becker:2006ks, Becker:2007dn, Becker:2007ee} and has been recently revisited and extended in \cite{Ishiguro:2021csu, Bardzell:2022jfh, Becker:2022hse, Cremonini:2023suw, Becker:2023rqi, Becker:2024ijy, Ishiguro:2024coq}. This allows one to explore flux compactifications away from the large volume, large complex structure, and even away from the weak coupling limit and thus provides a very interesting new place to test swampland conjectures.

While one can study these LG models in the large complex structure limit and find AdS vacua \cite{Becker:2006ks, Ishiguro:2021csu, Bardzell:2022jfh}, including infinite families, the detailed properties of these solutions are currently not under computational control unless one is also at weak coupling. However, for Minkowski vacua, one can study many properties of the vacua at strong coupling. This is due to the fact that string loop corrections only enter in the K\"ahler potential. The superpotential $W$ in these models is exact and does not receive perturbative or non-perturbative corrections \cite{Becker:2006ks}. This makes it, in particular, possible to calculate the number of massive fields in Minkowski vacua \cite{Becker:2022hse} as well as the number of massless fields that get stabilized via higher-order interactions \cite{Becker:2024ijy}. Such fields are similar to a massless scalar field $\phi$ with a $\phi^4$ interaction term. In our case, higher order terms in the superpotential can constrain the dimensionality of the solution space of the $\mathcal{N}=1$ Minkowski equations $W=\partial_i W = 0$ and can thereby stabilize massless fields.

These supersymmetric Minkowski vacua in the non-geometric LG models share an important feature with their geometric analogue vacua in GKP \cite{Giddings:2001yu}, namely the $G_3$ flux has to be imaginary-self-dual (ISD). This means that it contributes to the O3-plane tadpole with the same sign as D3-branes. The fixed number of O3-planes provides then an upper bound on the amount of flux that we can turn on. While early studies \cite{Giryavets:2003vd, Denef:2004dm, Denef:2005mm} seemed to confirm the naive expectation that generically any chosen ISD $G_3$ flux would stabilize all complex structure moduli, this has been called into question recently. For example, in the context of the sextic CY-fourfold, there is a tension between stabilizing all complex structure moduli and satisfying the tadpole cancellation condition \cite{Braun:2020jrx}. This and similar observations for other compactifications have led to the so-called Tadpole Conjecture \cite{Bena:2020xrh}.

The Tadpole Conjecture states that the fluxes used to stabilize complex structure moduli contribute to the tadpole an amount $N_{\rm flux}$ that grows unacceptably fast with the number of moduli that are massive (or get stabilized). In particular, it is conjectured to be true that\footnote{The first version of this paper and previous papers studying moduli stabilization in the $1^9$ LG model \cite{Becker:2022hse, Becker:2024ijy}, did not have the factor of 2 in the equation below. This factor of 2 arises from us defining the flux tadpole in equation \eqref{eq:tadpole} in the covering space following \cite{Becker:2006ks}. The original Tadpole Conjecture paper~\cite{Bena:2020xrh} has in equation (2.1) a factor of 1/2 in front of the flux contribution, effectively counting the flux contribution in the quotient space. We thank Daniel Junghans for bringing this to our attention.}
\be
N_{\rm flux} > 2\, \alpha\, n_{\rm stab}\,.
\ee
where the Refined Tadpole Conjecture states that $\alpha=1/3$.
For supersymmetric Minkowski vacua, this provides an upper bound on $n_{\rm stab}$, which is the number of moduli we can stabilize. This is due to the O3-plane tadpole cancellation condition
\be
N_{\rm flux} + N_{D3} = \frac{N_{O3}}{2}\,,
\ee
where $N_{O3}$ is the finite number of O3-planes and $N_{D3} \geq 0$ the number of D3-branes, both counted in the covering space.

The Tadpole Conjecture received a lot of attention recently and has been tested in many setups \cite{Bena:2021wyr, Marchesano:2021gyv, Plauschinn:2021hkp, Lust:2021xds, Bena:2021qty, Grana:2022dfw, Tsagkaris:2022apo, Becker:2022hse, Lust:2022mhk, Coudarchet:2023mmm, Braun:2023pzd, Braun:2023edp, Becker:2024ijy}. These tests have been performed in the asymptotics as well as the interior of moduli space. However, none of these papers found a violation of the Tadpole Conjecture except \cite{Becker:2022hse, Becker:2024ijy}. These two papers that studied the $1^9$ LG model, confirmed the linear relationship proposed by the Tadpole Conjecture, but found a violation of the Refined Tadpole Conjecture by almost a factor of 2. The $2^6$ LG model that we study in this paper allows for violations of the Refined Tadpole Conjecture by more than a factor of 2, as we will explain in this paper. This result was independently obtained in \cite{Becker:2024ayh}, where the authors initiate a full classification of all flux configurations in this model.

While the Tadpole Conjecture is mostly relevant for compactifications with a large number of complex structure moduli, it was independently conjectured that all 10d supergravity compactifications to 4d Minkowski solutions should always have a massless scalar field \cite{Andriot:2022yyj}. This so-called Massless Minkowski Conjecture is satisfied in all known supergravity compactifications but violated in this non-geometric $2^6$ LG model \cite{Becker:2024ayh}. The existence of 4d $\mathcal{N}=1$ vacua without massless fields is extremely interesting and shows how useful these non-geometric LG models are in improving our understanding of the string theory landscape of vacua. 

In this paper in section \ref{sec:review} we will review the $2^6$ LG model, mostly following \cite{Becker:2006ks}, adapted to the notations of \cite{Becker:2024ijy}. Then we will present a large number of different flux choices and calculate the number of massive fields and some higher-order terms to check for the stabilization of massless fields in section \ref{sec:higherorder}. We discuss our results in the context of moduli stabilization and the above-mentioned swampland conjectures in section \ref{sec:swamp}. We conclude by summarizing our findings in section \ref{sec:conclusion}.

\section{Review of the $2^6$ model}
\label{sec:review} 
The $2^6$ Gepner model admits a Landau-Ginzburg description and lives in the would-be K\"ahler moduli space of a rigid Calabi-Yau threefold with $90$ K\"ahler moduli. It is a tensor product of six minimal models, each with central charge $c=\frac 32$, and another factor theory which is trivial in the sense that it can be integrated out. The world-sheet superpotential of the tensor theory is
\be
\label{eq:WS-supo}
\cW = \sum_{i=1}^6 x_i^4 + z^2~,
\ee
and we orbifold by the $\bZ_4$ action freely generated by
\be 
\label{eq:Z4-action}
    g : x_i \mapsto \rmi x_i \quad \quad \quad z \mapsto -z~.
\ee
We will not integrate out the quadratic field $z$ since it would require the action of the orbifold on the chiral fields $x_i$ to be dressed by $(-1)^F$ which, in turn, would make the construction of A-branes in the orbifoldized theory cumbersome. In order to obtain 4d $\mathcal{N}=1$ supersymmetry, this theory should now be orientifolded. There are several choices of orientifold projections, as explained in \cite{Becker:2006ks}. In this paper we choose the canonical choice, denoted $\sigma_0$ in \cite{Becker:2006ks}, which acts as follows:
\be
\label{eq:sig0-Orientifold}
    \sigma_0: (x_1, \ldots, x_6) \mapsto \omega (x_1, \ldots x_6) \quad \quad \quad z \mapsto \rmi z
\ee
where $\omega\equiv e^{\frac{2 \pi \ii}{8}}$. It is easy to see that $\sigma_0^2 = g$.

The chiral (c) and anti-chiral (a) fields in the left and right moving sectors of this 2d, $\cN = (2,2)$ superconformal field theory, determined uniquely by the superpotential \eqref{eq:WS-supo}, possess a ring structure. These rings are the LG analogs of the cohomology rings of Calabi-Yau manifolds. Spectral flow relates the (anti-)chiral fields to Ramond ground states, and the $U(1)$-charges of the states determine the Hodge degrees of the harmonic forms \cite{Witten:1993yc,vafaOrbifoldized,Brunner:2004zd}. In particular, the (c,c) ring arises solely from the states in the untwisted sector of the Hilbert space. Concretely, it is the ring 
\be
\label{jacobi}
\cR = \biggl[\frac{\mathbb{C}\left[x_1, \ldots, x_6\right]}{\partial_{x_i} \cW\left(x_1, \ldots,
x_6\right)}\biggr]^{\mathbb{Z}_4}\,.
\ee
It is a $182$-dimensional complex vector space, spanned by monomials 
\begin{equation}
\label{eq:kk}
\xx^\kk = x_1^{k_1} \cdot x_2^{k_2} \cdots x_6^{k_6}
\end{equation}
with $\kk = (k_1,\ldots,k_6)$, $k_i\in \{0,1, 2\}$ for all $i$, and $\sum k_i = 0 \bmod 4$.
The elements with $\sum k_i=4$ are --- $15$ monomials of the form $x_i x_j x_k x_l$ with $i, j, k, l$ all distinct, $60$ monomials of the form $x_i^2 x_j x_k$ with $i, j, k$ all distinct, $15$ monomials of the form $x_i^2 x_j^2$ with $i, j$ distinct. These $90$ monomials span the space of marginal deformations of the theory. Concretely, introducing deformation parameters $t^\kk$, we write
\begin{equation}
\label{eq:W-deformation}
\cW\bigl(\{x_i\}\bigr) = \sum_{i=1}^{6}x_{i}^{4} \;\;\longrightarrow\;\; \cW\bigl(\{x_i\}; \{ t^\kk
\} \bigr) = \sum_{i=1}^6 x_i^4 - \sum_{\substack{\kk \\[.05cm]
\sum \!k_i=4}} t^\kk \xx^{\kk}~.
\end{equation}
The parameters $t^\kk$ are analogues of complex structure moduli. In addition, there are $3$ twisted sectors in the orbifold theory, twisted by $e^{-2 \pi \rmi \nu J_0}$, $\nu = 1,2,3$. The states in the $\nu=2$ sector do not survive the $U(1)$-projection, while the other two sectors contribute no non-trivial elements in the (a,c) ring. Therefore, there are no K\"ahler moduli in this model, making it non-geometric. The total number of moduli is $91$ --- $90$ complex structure moduli $t^\kk$, and the axio-dilaton $\tau = C_0 + \rmi\, e^{-\phi}$.

\subsection{Homology and Cohomology}
A-type D-branes in a Landau-Ginzburg model with one chiral field $x$ are contours in the $x$-space that map to a line of constant phase in the $\cW$-plane emanating from the critical value (assuming there is only one critical point, which is the case for the $2^6$ model) of $\cW$ \cite{HoriIqbalVafa}. Using an R-rotation, we look for preimages in the $x$-plane of the positive real line
\be
\Im \cW = 0
\ee
in the complex $\cW$ plane.
\begin{figure}[H]
    \centering
    \includegraphics[scale=0.435]{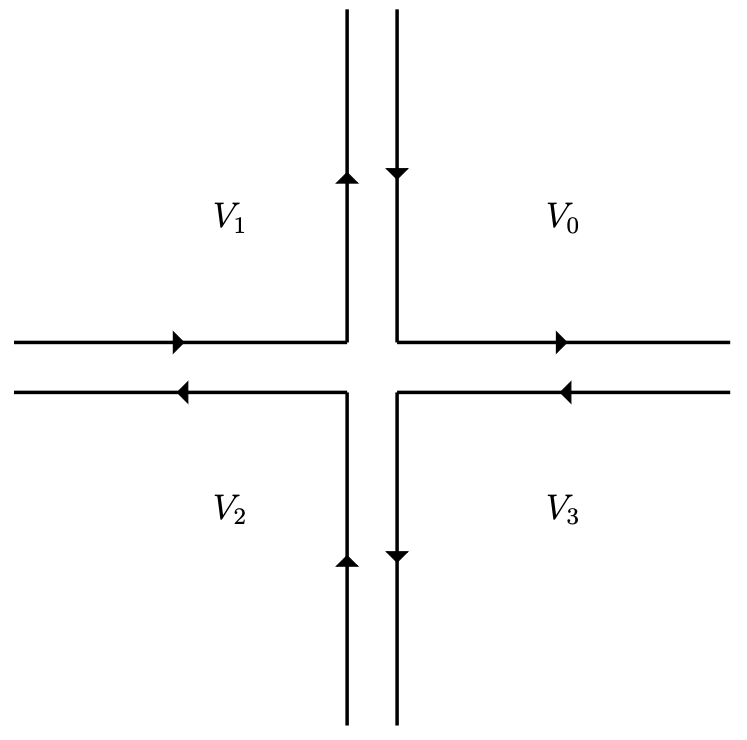}
    \caption{The four contours $\left(V_{0},V_{1},V_{2},V_{3}\right)$ in the complex $x$-plane.
    \label{fig:contours}}
\end{figure} 
For the minimal model of interest,
\be
\label{eq:WS-supo-factor}
    \cW = x^4~,
\ee
a homology basis is given by the four wedges $V_0$, $V_1$, $V_2$, and $V_3$ (see Fig. \ref{fig:contours}), which satisfy the relation 
\be
\label{eq:sumVn=0}
    V_0 + V_1 + V_2 + V_3 = 0~.
\ee
The action \eqref{eq:Z4-action} of $\bZ_4$ on $V_n$, $n=0,1,2,3$, is given by
\be
\label{eq:Z4-action-Vn}
    g: V_n \mapsto V_{(n+1) \bmod 4}~.
\ee
The matrix representation of \eqref{eq:Z4-action-Vn} is 
\be
\text{Mat}(g) = \begin{pmatrix}
0 & \hspace{0.3cm} 1 & \hspace{0.3cm} 0 & \hspace{0.3cm} 0 \\
0 & \hspace{0.3cm} 0 & \hspace{0.3cm} 1 & \hspace{0.3cm} 0 \\
0 & \hspace{0.3cm} 0 & \hspace{0.3cm} 0 & \hspace{0.3cm} 1 \\
1 & \hspace{0.3cm} 0 & \hspace{0.3cm} 0 & \hspace{0.3cm} 0
\end{pmatrix}
\ee
Following \cite{HoriIqbalVafa}, the intersection $\langle V_m | V_n \rangle $ can be geometrically computed as the geometric intersection between the two wedges, with $V_n$ held fixed and $V_m$ rotated counter-clockwise by a small angle:
\be
\label{eq:intersection-VmVn}
\bigl(\braket{V_{m}|V_n}\bigr)_{m,n=0,1,2,3} = \begin{pmatrix}
    1 & -1 & 0 & 0 \\ 
    0 & 1 & -1 & 0 \\ 
    0 & 0 & 1 & -1 \\
    -1 & 0 & 0 & 1
    \end{pmatrix} = \mathbf{I} - \textrm{Mat}(g)~.
\ee
The matrix elements can be recast as
\be
\label{eq:intersection-VmVn-matrix-elements}
    \langle V_m | V_n \rangle = \delta_{m,n} - \delta_{m+1, n}~,
\ee
where the Kronecker delta is understood modulo $4$ here and everywhere else below. The chiral ring of the minimal model \eqref{eq:WS-supo-factor} is $\cR = \bC[x]/x^3$ which, as a vector space, is the span of $1, x, x^2$. These basis elements correspond to the Ramond-Ramond ground states $|l\rangle$, $l=1,2,3$,
\be
\label{eq:flow}
    x^{k=0,1,2} \longleftrightarrow |l=1,2,3\rangle~.
\ee
Up to an overall factor, the states $|l\rangle$ will be the most useful ingredients in describing $3$-form fluxes in the orbifold theory. The non-geometric analogue of the integral of a $3$-form flux on a $3$-cycle is the pairing between the middle-dimensional homology and cohomology. To compute this, we need the overlap between the boundary states represented by the $V_n$ and the Ramond-Ramond ground states \cite{HoriIqbalVafa}:
\be
\label{eq:overlap-factor}
    \langle V_n | l \rangle = \int_{V_n} x^{l-1} e^{-\cW} = \frac{1}{4} \, \rmi^{nl} \, (1 - \rmi^l) \, \Gamma \Bigl( \frac{l}{4} \Bigr) ~.
\ee
Now we define the convenient ``$\Omega$-basis" \cite{Becker:2006ks, Becker:2023rqi, Becker:2024ijy} for the charge lattice, which enables fast computation of the derivatives of the spacetime superpotential with respect to moduli in the orbifoldized tensor theory. We choose the normalization
\be
\label{eq:factor-Omega-l}
    |\Omega_l \rangle := \frac{4}{\Gamma \Bigl(\frac{l}{4} \Bigr)} | l \rangle ~,
\ee
implying
\be
\label{eq:overlap-VnOmegal}
    \langle V_n | \Omega_l \rangle = \rmi^{nl} (1 - \rmi^l) ~.
\ee
Comparing \eqref{eq:intersection-VmVn-matrix-elements} and \eqref{eq:overlap-VnOmegal}, we have defined
\be
\label{eq:Omegal-Vn}
    |\Omega_l \rangle = \sum_n \rmi^{nl} |V_n \rangle~,
\ee
with the inverse relation
\be
\label{eq:Vn-Omegal}
    |V_n \rangle = \frac{1}{4} \sum_l \rmi^{-nl} | \Omega_l \rangle~.
\ee
For completeness, we record
\be
\label{eq:intersection-Om}
    \langle \Omega_{l'} | \Omega_l \rangle = 4 ~\delta_{l' + l, 4} ~ (1-\rmi^l)~.
\ee
Reality of $V_n$ and eq. \eqref{eq:Omegal-Vn} together imply the following conjugation rule on $\Omega_l$:
\be
\label{eq:Omega-bar}
    \overline{\Omega_l} := \Omega_{4-l} =: \Omega_{\bar l}~,
\ee
where we have defined the conjugate index notation $\bar l := 4-l $.
The astute reader may have noticed that the factor of $4$ is switched in equations \eqref{eq:Omegal-Vn} and \eqref{eq:Vn-Omegal} compared to the conventions in the recent paper \cite{Becker:2024ijy} on the $1^9$ LG model. The present choice is made following the conventions of\footnote{The concurrently published paper \cite{Becker:2024ayh} also adopts the same normalization as ours.} \cite{Becker:2006ks}. We will say more on this soon.

In the full tensor theory, the Ramond ground states are tensor products of the six factors, labeled by $|\ll \rangle$, $\ll = (l_1, l_2, \ldots, l_6)$, $l_i \in \{1,2,3\}$. The states that survive the orbifold projection \eqref{eq:Z4-action} are further restricted to have $\sum_i l_i = 6, 10, 14, 18$. Following \cite{Brunner:2004zd}, these can be Hodge decomposed into classes $H^{(p,q)}$, $p+q=3$, as shown in table~\ref{tab:Hodge}.
\begin{table}[h]
\centering
\begin{tabular}{|c|c|c|c|c|}
\hline$\sum_i l_i$ & 6 & 10 & 14 & 18 \\
\hline$H^{(p, q)}$ & $H^{(3,0)}$ & $H^{(2,1)}$ & $H^{(1,2)}$ & $H^{(0,3)}$ \\
\hline
\end{tabular}
\caption{Hodge decomposition of RR ground states in the $2^6$ LG model}
\label{tab:Hodge}
\end{table}

Similar to the states $|\ll \rangle$, the $3$-cycles are obtained by tensoring the wedges\footnote{For the factor theory $\cW = z^2$, there are two wedges that are overlapping straight lines oriented in opposite directions. We suppress these straight wedges in our notation for tensored cycles henceforth but carefully incorporate the sign-flip due to the $\bZ_4$ action on them.} $V_n$, and then taking a $\bZ$-span. Concretely, $V_\nn = V_{n_1} \times V_{n_2} \times \ldots \times V_{n_6}$, $\nn = (n_1, n_2, \ldots, n_6)$, $n_i \in \{0, 1, 2, 3\}$ for all $i=1,2,\ldots, 6$. The set $\{V_\nn\}$ is an over-complete basis due to the relation \eqref{eq:sumVn=0}. In the orbifoldized theory, an over-complete basis is induced by summing over $\bZ_4$ images:
\be
    |\gamma_\nn\rangle :=  |V_\nn\rangle - |V_{\mathbf{n+1}}\rangle + |V_{\mathbf{n+2}}\rangle - |V_{\mathbf{n+3}}\rangle ~.
\ee
Here and in what follows, $\mathbf{1} = (1,1,1,1,1,1)$, $\mathbf{2} = 2 \cdot \mathbf{1}$, and so on. The minus signs in the second and the fourth terms above are an artifact of the $\bZ_4$ action \eqref{eq:Z4-action} on $z$. Let us now count the number of independent cycles in the middle-dimensional homology. With $n_i \in \{0,1,2,3\}$, there are $4^6 = 4096$ different choices of $|\gamma_\nn\rangle$. This number gets reduced by a lot because of the relation \eqref{eq:sumVn=0} on each factor. Naively, there are ${6 \choose 1}\cdot 4^5 = 6144$ constraints if one imposes \eqref{eq:sumVn=0} factor-by-factor. These are the vectors $\sum_{n_i} |V_{(n_1 \ldots n_6)}\rangle$, $i=1,2,\ldots , 6$, set to zero. Clearly, since we have $4096$ cycles $|\gamma_\nn\rangle$ to begin with, many of these $6144$ constraints are linearly dependent. To correct for this, we reduce the number of constraints by ${6 \choose 2}\cdot4^4 = 3840$, which is the number of constraints satisfied by the constraints themselves\footnote{See Appendix C of \cite{Brunner:2004zd} for the same argument applied to the quintic.}. This, again, is an over-correction which we remedy by adding ${6 \choose 3}\cdot4^3 = 1280$. Continuing this way, it is easy to see that the number of unconstrained cycles is\footnote{A much faster way to arrive at the number $729$ is to restrict to $n_i \in \{0,1,2\}$ from the beginning, since $V_3 = - (V_0 + V_1 + V_2)$, and realize that we have $3^6 = 729$ choices for $\nn$. However, the effect of orbifolding is easier to work out from \eqref{eq:num-gamman-pre-orbifold}.}
\be
\label{eq:num-gamman-pre-orbifold}
    4^6 - {6 \choose 1}\cdot 4^5 + {6 \choose 2}\cdot 4^4 - \ldots + 1 = (4-1)^6 = 729~.
\ee
The above argument can be captured abstractly in terms of a long exact sequence as was done for the $1^9$ model in \cite{Becker:2024ijy} and for the quintic in \cite{Brunner:2004zd}. Upon orbifolding, identifications take place of cycles $|\gamma_\nn\rangle$ and, consequently, the constraints imposed on them. Each summand in the left-hand side of \eqref{eq:num-gamman-pre-orbifold} gets divided by a factor of $4$ except the last one which counts a $\bZ_4$-invariant constraint. Therefore, one gets $\frac{(4-1)^6 -1}4 +1 = 183$ independent cycles in the sixth tensor power of the minimal model $\cW = x^4$. Now including the $\cW=z^2$ factor, one can repeat the above counting exercise. We start with $183 \times 2$ cycles -- the factor of $2$ coming from the two wedges of the $z$-theory -- and we impose the constraint that the two straight wedges in the $z$-theory add up to zero. This yields $183\times2-1$ cycles. Orbifolding by the $\bZ_4$ action on $z$, this number reduces to $\frac{183 \times 2}{2} -1 = 182$. We will continue to use the same notation $|\gamma_\nn\rangle$, $\nn = (n_1, n_2, \ldots, n_6)$, for the cycles even after the inclusion of the $z$-factor. In summary, the rank of the lattice spanned by the $|\gamma_\nn\rangle$ in the full orbifoldized theory is $182$.

The overlap \eqref{eq:overlap-VnOmegal} generalizes to
\be
\label{eq:overlap-tensor-gamman-Oml}
    \langle \gamma_\nn | \Omega_\ll \rangle =  \rmi^{\nn . \ll} \prod_{i=1}^6 (1-\rmi^{l_i})~,
\ee
with $\nn . \ll := \sum_i n_i l_i$, where we have divided by a factor of $4$ due to orbifolding. The $\Omega$-basis in the tensor product is given by
\be
\label{eq:Om-tensor}
    |\Omega_\ll \rangle = \sum_{\nn} \rmi^{\nn . \ll} |V_\nn \rangle = \sum_{[\nn]} \rmi^{\nn . \ll} |\gamma_\nn\rangle~.
\ee
The intersection form in this basis is a generalization of \eqref{eq:intersection-Om}:
\be
\label{eq:intersection-Om-tensor}
    \langle \Omega_{\ll'} | \Omega_\ll \rangle = 4^5~ \delta_{\ll' + \ll , \mathbf{4}} \prod_{i=1}^6 (1-\rmi^{i_i})~,
\ee
where we have again divided by a factor of $4$ due to orbifolding. This is the normalization used in Sec. $5$ of \cite{Becker:2006ks}. Using the $|\gamma_\nn\rangle$, we have
\be
\label{eq:intersection-gamma-tensor}
    \langle \gamma_\nn' | \gamma_\nn \rangle = \langle V_{\nn'} | V_\nn \rangle - \langle V_{\nn' + \mathbf 1} | V_\nn \rangle + \langle V_{\nn' + \mathbf 2} |  V_\nn \rangle - \langle V_{\nn' + \mathbf 3} | V_\nn \rangle~.
\ee
The intersection matrix in the $|\gamma_\nn\rangle$-basis is most conveniently computed by first going to a truncated set and then carving out a full-rank sub-matrix, as described in \cite{Becker:2006ks}. We skip the details here but point out that this matrix has rank $182$, consistent with our discussion on the rank of the lattice spanned by the $|\gamma_\nn\rangle$.

\subsection{$3$-form fluxes}
Having set up the basic ingredients and normalization conventions, we can now quickly summarize the description of $3$-form fluxes in the LG language. Just as in the $1^9$ model \cite{Becker:2006ks, Becker:2024ijy}, the complexified $3$-form flux $G_3 = F_3 - \tau H_3$ can be expressed in two equivalent ways:
\be
\label{eq:G3-gamma-Omega}
    G_3 = \sum_\nn (N^\nn - \tau M^\nn) \, \gamma_\nn = \sum_\ll A^\ll \, \Omega_\ll~,
\ee
where $N^\nn, M^\nn \in \bZ$, and $A^\ll \in \bC$. Supersymmetry requires $G_3 \in H^{(2,1)} \oplus H^{(0,3)}$, with 
\be
    (G_3)_{\rm Mink} \in H^{(2,1)}~.
\ee
The expansion in the integral cohomology basis $\gamma_\nn$ is non-unique since $\{\gamma_\nn\}$ is over-complete to describe $H^{(2,1)} \oplus H^{(0,3)}$. The set of flux numbers $N^\nn, M^\nn \in \bZ$ is linearly dependent -- there are $182 \times 2 = 364$ of them, but their $\bZ$-span is a lattice of rank\footnote{The rank of the supersymmetric flux lattice can be derived by looking at the set of quantized ``1-$\Omega$" solutions as was done in \cite{Becker:2024ijy} for the $1^9$ model.} $182$. The expansion in the $\Omega$-basis, with complex coefficients $A^\ll$, is the most convenient for efficiently computing higher order terms of the spacetime superpotential. Moreover, by allowing $A^\ll$ to be non-zero only when $\sum_i l_i = 10$, we will restrict to $G_3 \in H^{(2,1)}$ leading to Minkowski solutions. Such fluxes span a sub-lattice of rank $180$. Dirac quantization of the fluxes reads
\be
\label{eq:fluxQuant}
    \int_{\gamma_\nn} G_3 = \langle \gamma_\nn | G_3 \rangle = N_\nn - \tau M_\nn~, ~~ N_\nn, M_\nn \in \bZ~.
\ee
Using \eqref{eq:overlap-tensor-gamman-Oml} and the $\Omega$-basis expansion in \eqref{eq:G3-gamma-Omega}, 
\be
    \langle \gamma_\nn | G_3 \rangle = \sum_\ll A^\ll \, \rmi^{\nn . \ll} \prod_{i=1}^6 (1-\rmi^{l_i}) = N_\nn - \tau M_\nn~.
\ee
The O-plane associated with the orientifold projection \eqref{eq:sig0-Orientifold} has charge $40$ that must be cancelled by the flux-tadpole $\tad$, and $N_{D3}$, the number of mobile D3-branes in the background. Using the overlap \eqref{eq:intersection-Om-tensor} in the $\Omega$-basis, we compute
\be\label{eq:tadpole}
    \tad = \frac{1}{\tau-\bar{\tau}} \langle G_{3} | \bar{G}_{3} \rangle = \frac{4^5}{\tau-\bar{\tau}} \sum_\ll |A^\ll|^2 \, \prod_{i=1}^6 (1- \rmi^{-l_i})~,
\ee
while the tadpole cancellation condition is
\be
    \tad \overset{!}{=} 40 - N_{D3}~.
\ee
The O-planes are able to contribute a much larger charge in the $2^6$ model to the tadpole than in the $1^9$ model. This is because the $1^9$ model corresponds to a toroidal orbifold with non-vanishing $B$ field, whereas, the $2^6$ model corresponds to a toroidal orbifold with vanishing $B$ field \cite{Becker:2006ks}. We choose $\tau = \rmi$ and, for $\Omega_\ll \in H^{(2,1)} \oplus H^{(0,3)}$, the product $\prod_{i=1}^6 (1- \rmi^{-l_i})$ is purely imaginary with a positive imaginary part. This confirms $\tad \geq 0$.

\subsection{The spacetime superpotential to all orders}
The Gukov-Vafa-Witten (GVW) superpotential $W$ \cite{Gukov:1999ya} engendered by a flux $G_3$ in geometric compactifications is given by 
\be
\label{eq:W-GVW}
    W = \int G_3 \wedge \Omega = \int (F_3 - \tau H_3) \wedge \Omega~,
\ee
where $\Omega$ is the holomorphic $3$-form in the internal space. This has an analogue in the LG language. Identifying $\Omega \in H^{(3,0)}$ with the RR ground state $|\11\rangle$ according to table~\ref{tab:Hodge}, we have for the $2^6$ LG model
\be
\label{superpotential}
    W = \langle G_3 | \11 \rangle = \sum_\ll A^\ll \, \langle \Omega_\ll | \11 \rangle~.
\ee
The overlaps in the factor theory, and consequently in the orbifoldized tensor theory, are deformed under the ``complex structure" deformations parameterized by\footnote{As per eqn. \eqref{eq:kk}, the vectors $\kk$ are in one-to-one correspondence with $\ll$ such that $\Omega_\ll \in H^{(2,1)}$ -- $\kk = \ll - \11$.} $t^\kk$. This is through the deformation of the worldsheet superpotential via \eqref{eq:W-deformation}. In a factor minimal model, $\cW \mapsto x^4 -tx$ leads to the contour integral \footnote{The contour integral has convergence issues unless $\Re t >0$. However, the derivatives are all well defined at $t=0$. This is consistent with the fact that we can only compute the superpotential and its derivatives at the Fermat point.}
\be
\label{eq:overlap-factor-deformed}
    \langle V_n | l \rangle_{\rm deformed} = \int_{V_n} x^{l-1} e^{-x^4 + t\,x} \, dx~.
\ee
The overlap \eqref{eq:overlap-factor-deformed} can be expanded in a Taylor series around $t=0$, with coefficients
\be
\label{eq:overlap-der}
    \Bigl(\frac{\partial}{\partial t}\Bigr)^r \Big|_{t=0} \langle V_n | l \rangle_{\rm deformed} = \int_{V_n} x^{r+l-1} \, e^{-x^4} \, dx = \frac{1}{4} \, \rmi^{n(r+l)} \, (1 - \rmi^{r+l}) \, \Gamma \Bigl( \frac{r+l}{4} \Bigr)~.
\ee
In the full theory, the dependence of $W$ on the deformation parameters $t^\kk$ introduced in \eqref{eq:W-deformation} is captured in a multi-variable Taylor expansion around $t^\kk = 0$, with coefficients that are derivatives of the overlaps $\langle \Omega_\ll | \11 \rangle$ in \eqref{superpotential} to be computed using \eqref{eq:overlap-der} as a starting point. Without belaboring this calculation, we quote the result:

\begin{equation}
\label{eq:Wdertt}
    \frac{\partial}{\partial t^{\mathbf{k}_1}} \frac{\partial}{\partial t^{\mathbf{k}_2}} \ldots \frac{\partial}{\partial t^{\mathbf{k}_r}} \bigg|_{\substack{t^\kk =0 \\[0.05cm]
    \tau = \tau_0}} W  = \sum_{\substack{\ll : \\[.05cm]
\bar \ll = \LL \bmod 4}} \frac 12 ~ A_\ll ~ \prod_{i=1}^6 (1 - \rmi^{L_i}) \Gamma \lp\frac{L_i}{4}\rp~,
\end{equation}
where we have defined
\begin{equation}
    \LL := \sum_{\alpha =1}^r \kk_\alpha + \mathbf{1}~.
\end{equation}
$W$ also depends linearly on the axio-dilaton $\tau$ through $G_3$. Therefore, we can also compute
\begin{align}
\label{eq:Wderttau}
    \frac{\partial}{\partial \tau} \frac{\partial}{\partial t^{\mathbf{k}_1}} \frac{\partial}{\partial t^{\mathbf{k}_2}} \ldots \frac{\partial}{\partial t^{\mathbf{k}_r}} \bigg|_{\substack{t^\kk =0 \\[0.05cm]
    \tau= \tau_0}} W  &= \sum_{\substack{\ll : \\[.05cm]
\bar \ll = \LL \bmod 4}} \frac {-\rmi}{4 ~ \rm{Im}(\tau_0)} ~ A_\ll ~ \prod_{i=1}^6 (1 - \rmi^{L_i}) \Gamma \lp\frac{L_i}{4}\rp~ \nonumber \\
    & ~ + \sum_{\substack{\ll : \\[.05cm]
 \ll = \LL \bmod 4}} \frac {\rmi}{4 ~ \rm{Im}(\tau_0)} ~ \overline{A_\ll} ~ \prod_{i=1}^6 (1 - \rmi^{L_i}) \Gamma \lp\frac{L_i}{4}\rp~,
\end{align}
while obviously
\begin{align}
    \Bigl(\frac{\partial}{\partial \tau}\Bigr)^{p\geq2} \frac{\partial}{\partial t^{\mathbf{k}_1}} \frac{\partial}{\partial t^{\mathbf{k}_2}} \ldots \frac{\partial}{\partial t^{\mathbf{k}_r}} \bigg|_{\substack{t^\kk =0 \\[0.05cm]
    \tau= \tau_0}} W &= 0~.
\end{align}
Using \eqref{eq:Wdertt} and \eqref{eq:Wderttau}, we have
\begin{align}
\label{eq:Wfull}
W=  \sum_{\bf{l}} \sum_{r=1}^\infty &\frac{1}{r!}\left(\sum_{\{t^{\mathbf{k}_\alpha}\} \text { with } 
\bar \ll = \LL} \prod_{i=1}^6  t^{\mathbf{k}_1} 
t^{\mathbf{k}_2} \ldots t^{\mathbf{k}_r}\, ~ A_\ll ~ \prod_{i=1}^6 (1 - \rmi^{L_i})  \Gamma \lp\frac{L_i}{4}\rp \lp \frac 12 -\rmi \frac{\tau-\tau_0}{4 \Im(\tau_0)} \rp\right.\cr
&\left.+\rmi\sum_{\{t^{\mathbf{k}_\alpha}\} \text { with }  \ll = \LL } \prod_{i=1}^6 
t^{\mathbf{k}_1} t^{\mathbf{k}_2} \ldots t^{\mathbf{k}_r}\, 
\overline{A_\ll} ~ \prod_{i=1}^6 (1 - \rmi^{L_i}) \Gamma \lp\frac{L_i}{4}\rp\, \frac{\tau-\tau_0}{{4 \Im(\tau_0)}}\right)\,,
\end{align}
where we will set $\tau_{0}=\rmi$ and we refer to $\tau-\tau_{0}$ as $t^{0}$\,.

\section{Moduli stabilization at higher order}
\label{sec:higherorder} 
In this section, we examine the stabilization of massless fields via the higher-order terms in the superpotential \cite{Becker:2022hse}. Such an analysis was performed recently in \cite{Becker:2024ijy} for many different flux configurations in the $1^9$ LG model. The authors of that paper studied higher-order terms in the superpotential up to 6th order in the moduli. They found that, for some flux configurations, no massless fields get stabilized via higher-order terms although all moduli do appear in the superpotential. For other flux choices, more and more fields get stabilized at higher order but the number of newly stabilized massless fields seems to quickly tend to zero (see table 2 on page 35 of \cite{Becker:2024ijy}). Here we perform a similar analysis for the $2^6$ model, which produces new and surprising results.

\subsection{Review of higher-order stabilization}
To illustrate the algorithm, which is explained in detail in \cite{Becker:2024ijy}, we study a toy model. Consider the following superpotential
\be \label{eq:example}
W = \frac{1}{2}\left(\phi-\psi^2\right)^{2}\,,
\ee
where $\phi$ and $\psi$ are complex fields. It necessarily vanishes at its critical points and its vanishing locus is given by $\phi = \psi^2$. We expand this superpotential order-by-order around a critical point to arrive at the same conclusion. Whilst this is redundant for a function as simple as the one in $\eqref{eq:example}$, it will be relevant for our analysis later as we will only be able to find supersymmetric Minkowski critical points of the GVW superpotential and compute its derivatives at a specific point in the moduli space. The origin, $\left(\phi,\psi\right)=\left(0,0\right)$ is clearly a critical point of the superpotential where it vanishes. As a result, expanding around the origin we have,
\be \label{eq:expand1}
W= \frac{1}{2!} \partial_i \partial_j W|_{\{t^{i}\}=0} t^{i} t^{j} + \frac{1}{3!} \partial_i \partial_j \partial_kW|_{\{t^{i}\}=0} t^{i} t^{j} t^{k} + \frac{1}{4!} \partial_i \partial_j \partial_k \partial_l W|_{\{t^{i}\}=0} t^{i} t^{j} t^{k} t^{l}\,,
\ee
where $t^{i} \in \{\phi, \psi\}$. The series truncates in this simple example but for the supersymmetric Minkowski critical points that we will study, the superpotential does not truncate when expanding around the critical points. As a result, schematically we will always have
\be \label{eq:WExpand}
W = \sum_{n\geq2} W_{n}\,,
\ee
where $W_{n}=\frac{1}{n!}(\partial_{i_{1}\ldots i_{n}}W) t^{i_{1}}\ldots t^{i_{n}}$. We would first like to ascertain the rank of the deformation space of the critical point after including terms up to quadratic order in the fields in $W$. This is done by studying the critical point equations,
\be \label{eq:critquad}
\begin{aligned}
\partial_{\phi}W_{2} &= \phi =0 \\
\partial_{\psi}W_{2} &= 0\,.
\end{aligned}
\ee
At this quadratic order, we find that one field $\left(\phi\right)$ is fixed up to linear order in the unfixed field $\left(\psi\right)$. Although it is not possible to determine which field directions are fixed without knowing the canonically normalized kinetic terms, we can count the number of fields that are fixed. The number of fields that are fixed at the quadratic order is equal to the rank of the Hessian of the superpotential \cite{Becker:2022hse} and so we will refer to these fields as being massive. The field $\phi$ is fixed up to linear order in the massless fields due to \eqref{eq:critquad} but could receive higher order corrections. The more precise statement would hence be that the critical point equations are evaluated as order-by-order expansions of the stabilized fields in terms of the unfixed fields, 
\be \label{eq:phiexpand}
\phi= \phi_{(1)}+\phi_{(2)}+ \phi_{(3)}+\ldots\,,
\ee
and we have determined that $\phi_{(1)}$ vanishes. The higher-order terms will get fixed sequentially as we include higher-order contributions in the superpotential. Now we wish to inspect whether the massless field gets lifted as we include higher-than-quadratic terms in $W$. First, we include cubic terms and analyze the critical point equations after truncating to quadratic order in the unfixed fields,
\be \label{eq:critcub}
\begin{aligned}
\partial_{\phi}W_{2}+\partial_{\phi} W_{3} &=   \phi_{(1)}+ \phi_{(2)} - \psi^2 =0 \\
\partial_{\psi}W_{2}+\partial_{\psi} W_{3} &= -2 \phi_{(1)} \psi= 0
\end{aligned}
\ee
Since $\phi_{(1)}=0$\,, the first equation fixes $\phi_{(2)}=\psi^2$\, and the second equation trivially vanishes. The field $\psi$ remains unfixed. After including quartic terms in the superpotential and truncating the critical point equations to cubic order in the unfixed fields we get
\be \label{eq:crtiquart}
\begin{aligned}
\partial_{\phi}W_{2}+\partial_{\phi} W_{3}+\partial_{\phi} W_{4} &=   \phi_{(1)}+ \phi_{(2)}- \psi^2+ \phi_{(3)} =0 \\
\partial_{\psi}W_{2}+\partial_{\psi} W_{3}+\partial_{\psi} W_{4} &=-2  \phi_{(1)} \psi-2  \phi_{(2)} \psi + 2  \psi^{3}=0\,.
\end{aligned}
\ee
Once we evaluate these equations at $\phi_{(1)}=0\,,\phi_{(2)}=\psi^2$\,, we find $\phi_{(3)}=0$\,. Since the series truncates at this order and the field $\psi$ is unfixed, we conclude that there is one flat direction and the origin is not an isolated critical point. 

The critical points of the superpotential in equation \eqref{eq:Wfull} that we will study have some similarities to the example we discussed above but in general will be more complicated. As was explained in \cite{Becker:2024ijy}, this algorithm has some limitations. For instance, since the series expansion around the critical point does not truncate in general, we will only be able to make definitive statements about the number of stabilized fields at a given order. The fields that remain unfixed are not necessarily flat directions and may be lifted at some higher order that we have not analyzed due to computational constraints. Another potential limitation is the appearance of branches. When solving the polynomial equations that arise as non-trivial critical point equations we find multiple branches of solutions which can lead to different numbers of stabilized fields. For example, consider the two equations
\be \label{eq:branches}
x y =0\,, \quad \quad x z=0\,.
\ee
They are solved by either $x=0$ or $y=z=0$. We will see explicit examples where this occurs in section \ref{subsec:branches}. The occurrence of these branches of solutions provides a challenge to single out the exact number of fields stabilized up to a given order in the superpotential. As a remedy to this issue, it was proposed in \cite{Becker:2024ijy} that the minimum number of fields stabilized across the different branches be taken as the number of stabilized fields at the critical point. For a more thorough mathematical discussion in terms of the Zariski and Krull co-dimensions of the critical points, we refer the reader to \cite{Becker:2024ijy}.

\subsection{One to three $\Omega$ flux choices}
In this subsection we study the simplest flux configurations, i.e., we turn on only 1, 2, or 3 different flux components in the $\Omega$-basis. While this makes some things very easy, it also has severe drawbacks as we describe below. We do not only consider physical solutions with $N_{\rm flux} \leq 40$ but also two unphysical ones.

There are three $S_{6}$ distinct single $\Omega$ solutions whose shortest, properly quantized, flux vectors are
\ba \label{eq:SingleOmega1}
G^{1} &= \frac{1}{32} \, \Omega_{1,1,2,2,2,2}\,,\\
\label{SingleOmega2}
G^{2} &= \frac{1}{16} \, \Omega_{1,1,1,2,2,3}\,,\\
\label{SingleOmega3}
G^{3} &= \frac{1}{8} \, \Omega_{1,1,1,1,3,3}\,.
\ea
The last solution $G^{3}$ has $N_{\rm flux}=64$ and it is thus unphysical. However, these three solutions (and their $S_6$ permutations) are the building blocks for all other solutions so we analyzed all three. 

For the first flux choice $G^1$ which has $N_{\rm flux}=16$ we find 36 massive fields and no higher order stabilization even when going to the 6th power in the superpotential. For the second flux choice $G^2$ with $N_{\rm flux}=32$ we actually find higher order stabilization already when including the cubic terms in the superpotential. However, we get 34 non-trivial quadratic constraints involving 34 massless fields. Unfortunately, we have not been able to solve these equations and therefore we do not know how many of the massless fields actually get stabilized at 3rd power in the superpotential. This is a recurring problem for these simple flux configurations with a small number of $\Omega$ components turned on. If they stabilize fields at higher orders, then there are many corresponding constraint equations that involve a large number of fields and we cannot solve them using standard techniques in, for example, Mathematica. Solution $G^3$ has $N_{\rm flux}=64$ and is therefore unphysical. It has 21 massive fields and 52 non-trivial quadratic constraints in 52 variables when including cubic terms in the superpotential. We are again unable to solve these. It would be interesting to understand why for certain flux configurations we find no moduli stabilization \cite{Grimm:2024fip}, but for now we simply present it here as an observation.

We studied the following flux choices with two $\Omega$ components and found similar results
\ba \label{eq:TwoOmega1}
G^{4} &=&  \left(-\frac{1}{64}+ \frac{\rmi}{64}\right) \Omega_{1,1,2,2,2,2} +\left(\frac{1}{64}+ \frac{\rmi}{64}\right) \Omega_{2,2,1,1,2,2} \,,\\
\label{eq:TwoOmega2}
G^{5}&=& \left(-\frac{1}{64}+ \frac{\rmi}{64}\right) \Omega_{1,1,2,2,2,2} -\left(\frac{1}{64} - \frac{\rmi}{64}\right) \Omega_{1,2,1,2,2,2} \,,\\
\label{eq:TwoOmega3}
G^{6} &=& \frac{1}{16} \, \Omega_{1,1,1,1,3,3} - \frac{1}{16} \, \Omega_{1,1,1,3,1,3}\,,\\
\label{eq:TwoOmega4}
G^{7} &=& \frac{1}{16} \, \Omega_{1,1,1,1,3,3} - \frac{1}{16} \,\Omega_{1,1,3,3,1,1}\,,\\ 
\label{eq:TwoOmega5}
G^{8} &=& \frac{1}{32} \, \Omega_{1,1,1,2,2,3} + \frac{1}{32} \, \Omega_{1,1,3,2,2,1}\,,\\
\label{eq:TwoOmega6}
G^{9} &=& \left(- \frac{1}{16} + \frac{\rmi}{16} \right) \Omega_{1,1,1,1,3,3} - \left(\frac{1}{64}-\frac{\rmi}{64}\right) \Omega_{1,1,2,2,2,2}\,,\\
\label{eq:TwoOmega7}
G^{10} &=& \left(-\frac{1}{16}+ \frac{\rmi}{16} \right) \Omega_{1,1,1,1,3,3} - \left(\frac{1}{32}-\frac{\rmi}{32} \right) \Omega_{1,1,1,2,2,3}\,,\\
\label{eq:TwoOmega8}
G^{11} &=& \left(-\frac{1}{16}+ \frac{\rmi}{16} \right) \Omega_{1,1,1,1,3,3} - \left(\frac{1}{32}- \frac{\rmi}{32} \right) \Omega_{1,1,2,3,1,2}\,.
\ea
A summary of the results for all the above flux choices and the one three $\Omega$ flux configuration  
\be
G^{12} = \left(\frac{1}{32}- \frac{\rmi}{32} \right) \Omega_{1,1,1,1,3,3} + \frac{1}{64} \Omega_{1,2,1,2,2,2} + \left(\frac{1}{32}+\frac{\rmi}{32} \right) \Omega_{1,3,1,3,1,1}
\ee
is presented in table \ref{tab:stabilization1to3}. \\

\begin{table}[H]
\begin{center}
\begin{tabular}{|c|c|c|c|c|c|c|
} 
 \hline
Model & $N_{\rm flux}$ & massive & 3rd power & 4th power & 5th power & 6th power 
\\ [0.5ex] 
 \hline
 $G^{1}$ & 16 & 36 & 0 & 0 & 0 & 0
 \\ 
 \hline
 $G^{2}$ & 32 & 28 & $\neq 0$ &  &  &
 \\
 \hline
 $G^{3}$ & 64  & 21 & $\neq 0$ &  &  &
 \\
 \hline
 $G^{4}$ & 16  & 52 & 0 & 0 & 0 & 0
 \\
 \hline
 $G^{5}$ & 16 & 36  & 0 & 0 & 0 & 0
 \\ 
 \hline
 $G^{6}$ & 32 & 28 & $\neq 0$ &  &  & 
 \\ 
 \hline
 $G^{7}$ & 32  & 36 & $\neq0$ &  &  & 
 \\ 
  \hline
 $G^{8}$ & 16 & 36 &  0 & 0 & 0 & 0
  \\
 \hline
  $G^{9}$ & 40 & 44  & $\neq 0$ &  &  & 
 \\ 
 \hline
 $G^{10}$ & 48 & 28 & $\neq 0$ &  &  & 
 \\ 
 \hline
 $G^{11}$ & 48 & 41 & $\neq 0$ &  &  & 
 \\ 
 \hline
  $G^{12}$ & 20 & 48 & 0 & 0 & 0 & 0
 \\ 
 \hline
\end{tabular}
\end{center}
\caption{A summary of the different most simple flux choices that we have analyzed. The second column lists $N_{\rm flux}$, the tadpole contribution of the particular flux configuration. The third column lists the number of massive fields and the following columns the number of fields that are massless but do get fixed due to terms of the $r$-th power in the superpotential that are polynomials. Whenever fields get stabilized we were not able to solve the corresponding constraints and could not pursue that flux choice to higher order. \label{tab:stabilization1to3}}
\end{table}

The clear result of the table is that solutions with $N_{\rm flux}=16$, which is really low, seem to have no further stabilization at higher order. However, solutions with $N_{\rm flux}\geq32$ generically lead to the stabilization of massless fields through cubic interactions. For that reason, we have included here the 3 $\Omega$ solution $G^{12}$ with $N_{\rm flux}=20$ to check at which point higher order stabilization takes place. We see that for this particular three $\Omega$ solution there are still no non-trivial higher-order constraints that stabilize massless fields. However, as we show in the next subsection, for a particular four $\Omega$ solution with $N_{\rm flux}=28$ there are higher order constraints that we can actually analyze.

Before ending this subsection we state the important result that the flux choice $G^4$ has $N_{\rm flux}=16$ and 52 massive fields. To our knowledge, this is the first instance that violates the Refined Tadpole Conjecture bound \cite{Bena:2020xrh} by more than a factor of 2: $N_{\rm flux}>\tfrac23 n_{\rm stab}$.\footnote{The tadpole condition in this example is not satisfied and requires the presence of 24 D3-branes. Those might have moduli associated with themselves but that is irrelevant for the fact that 52 complex structure moduli are massive for this simple flux choice.} We comment on this further in section \ref{sec:swamp} below.

\subsection{A four $\Omega$ flux choice} \label{subsec:branches}
Given the slightly disappointing outcome for the stabilization at higher order for the simplest solutions in the $2^6$ model, we now turn to an intermediate solution that is much more interesting. This flux choice was discovered by the authors of \cite{Becker:2024ayh} and they kindly allowed us to study the higher-order stabilization for it, already before they published their paper. The flux choice is 
\be \label{4Omega}
G^{13} = \left(\frac{1}{32}+\frac{\rmi}{32}\right) \Omega_{1,1,1,1,3,3}+\left(\frac{1}{32}+\frac{\rmi}{32}\right) \Omega_{1,3,1,3,1,1}+ \frac{1}{32}\Omega_{3,2,2,1,1,1}+\frac{\rmi}{64} \Omega_{1,1,2,2,2,2}
\ee
and it leads to $N_{\rm flux}=28$ and $75$ massive fields. After including cubic terms in the superpotential we find the following non-trivial quadratic constraints, 
\begin{align}
   t^{34} t^{87} &= 0  \cr
   t^{33} t^{86} &= 0 \cr
   t^{18} t^{52} &= 0   \cr
   t^{17} t^{51} &= 0  \cr
   t^{33} t^{51}  &= 0 \cr
   t^{34} t^{52}  &= 0 \cr
   t^{18} t^{51} + t^{17} t^{52} &= 0 \cr
   t^{50} t^{51} -(1-\rmi) t^{52} t^{58} &= 0 \\
   t^{48} t^{52} - (1-\rmi) t^{51} t^{58} &= 0 \cr
   t^{33} t^{38} - (1-\rmi) t^{34} t^{44}  &= 0 \cr
   t^{34} t^{36} - (1-\rmi) t^{33} t^{44}  &= 0 \cr
   t^{51} t^{52} + 2 \rmi t^{34} t^{86} + 2 \rmi t^{33} t^{87} &= 0 \cr
   t^{51} t^{62} + t^{38} t^{86} - (1-\rmi) t^{44} t^{87} &= 0 \cr
   t^{52} t^{64} + t^{36} t^{87} - (1-\rmi) t^{44} t^{86} &= 0 \cr
   t^{44} t^{52} + (1-\rmi)t^{17} t^{50} + (1-\rmi) t^{33} t^{62}+ 2 \rmi t^{18} t^{58}  &= 0 \cr
   t^{44} t^{51} + (1-\rmi)t^{18} t^{48} + (1-\rmi) t^{34} t^{64}+ 2 \rmi t^{17} t^{58}  &= 0\nonumber
\end{align}
Solving the above constraints leads to multiple branches of solutions. There are branches which fix $8$, $9$, $10$, and $12$ fields. The branches that stabilize $9, 10$ and, $12$ fields already violate the bound of the Refined Tadpole Conjecture by more than a factor of 2, which did not seem possible in the $1^9$ LG model \cite{Becker:2024ijy}. However, the branches that stabilize $8$ fields do not violate the bound by that much. As explained in \cite{Becker:2024ijy}, when such branches with different numbers of stabilized fields appear, we pick the branch with the least number of stabilized fields. So at cubic order, we do not have a violation of the bound by a factor of 2 yet. We then analyse the non-trivial cubic constraints arising after the inclusion of quartic terms in the superpotential. The minimum number of stabilized fields occur on branches that had $8$ or $9$ fields stabilized at cubic order. There are branches that had stabilized $9$ fields at cubic order, that stabilize no fields at quartic or quintic order, leaving us with $75+9+0+0=84=28\cdot3$ stabilized fields. So, this would violate the original Refined Tadpole Conjecture by a factor of 2. All branches that stabilized $8$ fields at cubic order stabilize at least $1$ field at quartic order, thus leading to $75+8+1=84=28\cdot3$ and the same conclusion as above. 

\begin{table}[H]
\begin{center}
\begin{tabular}{|c|c|c|c|c|
} 
 \hline
Model & $N_{\rm flux}$ & massive & 3rd power & 4th power 
\\ [0.5ex] 
 \hline
  $G^{13}$ & 28 & 75 & 8 & 1 
  \\ 
 \hline
  $G^{13}$ & 28 & 75 & 8 & 2 
 \\ 
 \hline
  $G^{13}$ & 28 & 75 & 8 & 3 
 \\ 
 \hline
  $G^{13}$ & 28 & 75 & 9 & 0 
 \\ 
 \hline
  $G^{13}$ & 28 & 75 & 9 & 1 
 \\ 
 \hline
  $G^{13}$ & 28 & 75 & 9 & 2 
 \\
 \hline
  $G^{13}$ & 28 & 75 & 9 & 3 
 \\ 
 \hline
  $G^{13}$ & 28 & 75 & 10 & 0 
 \\ 
 \hline
  $G^{13}$ & 28 & 75 & 10 & 1 
 \\ 
 \hline
  $G^{13}$ & 28 & 75 & 10 & 2 
 \\ 
 \hline
  $G^{13}$ & 28 & 75 & 12 & 0 
 \\ 
  \hline
\end{tabular}
\end{center}
\caption{A summary of the different branches for the flux choice $G^{13}$. The second column lists $N_{\rm flux}$ the tadpole contribution of this particular flux configuration. The third column lists the number of massive fields and the following columns the number of fields that are massless but do get fixed due to terms of the $r$-th power in the superpotential that are polynomials. We find different components that either fix the same or different numbers of fields, as indicated in the multiple rows for this single model. \label{tab:stabilization}}
\end{table}

\subsection{Close to Full Rank Solutions}
In this section, we present our study of 175 different solutions with a tadpole contribution of $N_{\rm flux}=39$ or $N_{\rm flux}=40$. They all have close to full mass matrix rank with between 86 and 90 massive fields. Since all of these solutions have many different $\Omega$ components with non-vanishing coefficients, it is not possible to present them all here and we have thus enclosed them in a Mathematica notebook that we attached as an ancillary file to this paper. 

For the curious reader, let us briefly mention how we obtained these high rank solutions. There is a straightforward algorithm to obtain an integral basis of the supersymmetric flux lattice in the present model. It is the same algorithm that was used in \cite{Becker:2023rqi} to derive an integral basis for Minkowski fluxes in the $1^9$ model. See also \cite{Becker:2024ijy} for a detailed description. We employed this algorithm in the $2^6$ model to obtain $180$ basis elements for the Minkowski lattice. Using ``small linear combinations" -- with four or five basis vectors linearly combined with small integer coefficients (between $-2$ and $2$) -- it was possible to generate solutions with a large variety of $(\tad, n_{\rm massive})$ values. All of our high rank solutions were obtained this way, with a remarkably small computational effort. The authors of \cite{Becker:2024ayh}, in their attempt to extensively test the tadpole conjecture, use this as one of the methods to generate a large number of solutions.

For all of these high rank flux configurations, we have checked the potential stabilization of massless fields via cubic and quartic terms in the superpotential. For several, there seems to be no further stabilization, but we also find several models where 1, 2, 3, or 4 fields get stabilized either through cubic, quartic, or quintic terms in the superpotential. Surprisingly, we have never found an example where fields get stabilized at cubic and quartic levels, contrary to what was found for the $1^9$ model in \cite{Becker:2024ijy}. 

For concreteness, we present here one example of a flux configuration with $N_{\rm flux}= 40$ that has 87 massive fields
\begin{align} \label{eq:BigOmega1}
G^{14} &= \frac{1}{32} \left[ -\frac{1}{2} \Omega _{1,1,1,1,3,3}+\frac{1}{2} \Omega_{1,1,3,3,1,1}-\frac{\rmi}{2} \Omega_{1,3,1,3,1,1} \right.\cr
&\qquad +\left(\frac{1}{4}+\frac{\rmi}{4}\right) \Omega_{1,3,2,1,1,2}+\left(\frac{1}{4}+\frac{\rmi}{4}\right) \Omega_{1,3,2,1,2,1}-\frac{1}{2} \Omega_{1,3,2,2,1,1} \cr
&\qquad-\frac{\rmi}{2} \Omega_{1,3,3,1,1,1}+\left(\frac{1}{4}+\frac{\rmi}{4}\right) \Omega_{2,1,1,1,2,3}+\left(\frac{1}{4}+\frac{\rmi}{4}\right) \Omega_{2,1,1,1,3,2}\cr
&\qquad-\left(\frac{1}{4}+\frac{\rmi}{4}\right) \Omega_{2,1,2,3,1,1}-\left(\frac{1}{4}+\frac{\rmi}{4}\right) \Omega_{2,1,3,2,1,1}-\left(\frac{1}{4}-\frac{\rmi}{4}\right) \Omega_{2,2,1,3,1,1}\cr
&\qquad -\frac{\rmi }{4} \Omega_{2,2,2,1,1,2}-\frac{\rmi }{4} \Omega_{2,2,2,1,2,1}+\left(\frac{1}{4}+\frac{\rmi}{4}\right) \Omega_{2,2,2,2,1,1}\\
&\qquad -\left(\frac{1}{4}-\frac{\rmi}{4}\right) \Omega_{2,2,3,1,1,1}-\left(\frac{1}{4}+\frac{\rmi}{4}\right) \Omega_{2,3,1,1,1,2}-\left(\frac{1}{4}+\frac{\rmi}{4}\right) \Omega_{2,3,1,1,2,1}\cr
&\qquad+\left(\frac{1}{4}+\frac{\rmi}{4}\right) \Omega_{2,3,1,2,1,1}+\left(\frac{1}{4}+\frac{\rmi}{4}\right) \Omega_{2,3,2,1,1,1}-\frac{\rmi}{2} \Omega_{3,1,1,1,1,3}\cr
&\qquad-\frac{\rmi }{2} \Omega_{3,1,1,1,2,2}-\frac{\rmi }{2} \Omega_{3,1,1,1,3,1}+\left(\frac{1}{2}+\frac{\rmi}{2}\right) \Omega_{3,1,1,3,1,1}\cr
&\qquad-\left(\frac{1}{4}-\frac{\rmi}{4}\right) \Omega_{3,1,2,1,1,2}-\left(\frac{1}{4}-\frac{\rmi}{4}\right) \Omega_{3,1,2,1,2,1}+\left(\frac{1}{2}+\frac{\rmi}{2}\right) \Omega_{3,1,3,1,1,1}\cr
&\qquad \left. +\frac{\rmi }{2} \Omega_{3,2,1,1,1,2}+\frac{\rmi }{2} \Omega_{3,2,1,1,2,1}-\frac{\rmi }{2} \Omega_{3,2,1,2,1,1}-\frac{\rmi }{2} \Omega_{3,2,2,1,1,1} \right]\,.\nonumber
\end{align}
The above flux choice has 32 $\Omega$ components turned on and satisfies the tadpole cancellation condition without the requirement of adding D3-branes. It has close to full mass matrix rank, which means that the 87 massive fields solve 87 of the 91 equations $\partial_i W=0$. After solving these 87 equations for the massive fields, we can plug the answer into the remaining four equations that are then functions of the four unstabilized fields only, which makes them easy to solve. When calculating $\partial_i (W_2+W_3)=0$ we find that these four equations are trivially satisfied. For $\partial_i (W_2+W_3+W_4)=0$ we find that 3 massless fields get stabilized and there are no further constraints when including $W_5$. Thus this model has, up to including $W_5$, a total of 87+0+3+0=90 stabilized fields. One might be encouraged by this and look for other examples that get to 91 or try to push this model to higher order in the fields. We did the former and studied 140 different flux configurations with $N_{\rm flux}=40$ and found a variety of different stabilization patterns at higher order as is summarized in table \ref{tab:N40}. In this table \ref{tab:N40} we did not pursue all different branches for all of these flux choices but only the one that stabilizes the least number of massive fields. \\

\begin{table}[H]
\begin{center}
\begin{tabular}{|c|c|c|c|c|c|
} 
 \hline
Multiplicity & $N_{\rm flux}$ & massive & 3rd power & 4th power & 5th power 
\\ [0.5ex] 
 \hline
 78 & 40 & 90 & 0 & 0 & 0 
 \\ 
 \hline
 1 & 40 & 89 & 1 & 0 & 0
 \\
 \hline
 2 & 40  & 89 & 0 & 1 & 0 
 \\
 \hline
 13 & 40 & 89  & 0 & 0 & 0
 \\ 
 \hline
 1 & 40 & 88 & 2 & 0 & 0 
 \\ 
 \hline
 5 & 40  & 88 & 0 & 1 & 0
 \\ 
  \hline
 2  & 40 & 88 &  0 & 0 & 1
  \\
  \hline
 5  & 40 & 88 &  0 & 0 & 0
  \\
 \hline
 1  & 40 & 87 & 3 & 0 & 0 
 \\ 
 \hline
 2 & 40 & 87 & 0 & 3 & 0 
 \\ 
 \hline
 8 & 40 & 87 & 0 & 2 & 0 
 \\ 
 \hline
 1 & 40 & 87 & 1 & 0 & 1 
 \\ 
 \hline
 5 & 40 & 87 & 0 & 1 & 0
 \\ 
 \hline
 2 & 40 & 86 & 0 & 4 & 0
 \\ 
 \hline
 4 & 40 & 86 & 0 & 2 & 0
 \\ 
 \hline
 3 & 40 & 86 & 0 & 1 & 0
 \\ 
 \hline
 7 & 40 & 86 & 0 & 0 & 0
 \\ 
 \hline
\end{tabular}
\end{center}
\caption{A summary of 140 different flux choices with $N_{\rm flux}=40$ that we have analyzed. The first column lists the number of $S_6$ distinct solutions that exhibit the shown pattern of massive fields and massless stabilized fields at the $r$-th power in the superpotential. It is of course possible that these different flux choices behave differently and split further if one were to go to even higher powers in the superpotential. \label{tab:N40}}
\end{table}

The flux choices in table \ref{tab:N40} all have $N_{\rm flux}=40$ and a mass matrix rank between 86-90 but are otherwise distinct and in a certain sense generated randomly. We might thus be tempted to do some statistical interpretation of our results. The table indicates that the stabilization of massless fields at higher order is less likely than no stabilization (at least to the order we work). Stabilizing fewer fields seems also more likely than stabilizing more fields. However, probably the most surprising result is that higher-order stabilization for all mass matrix ranks from 86-90 seems to not lead to all fields being stabilized. The maximum number of massless, stabilized field is given by $90$ minus the entry in the ``massive" column.

We have also generated a set of 33 solutions with $N_{\rm flux}=39$ and mass matrix ranks 86, 87, 88, and 89. We again studied higher-order stabilization and our results are summarized in table \ref{tab:N39}. In this table \ref{tab:N39} we did not pursue all different branches for all of these flux choices but only the one that stabilizes the least number of massless fields.

\begin{table}[H]
\begin{center}
\begin{tabular}{|c|c|c|c|c|c|
} 
 \hline
Multiplicity & $N_{\rm flux}$ & massive & 3rd power & 4th power & 5th power 
\\ [0.5ex] 
 \hline
 2 & 39 & 89 & 0 & 1 & 0 
 \\ 
 \hline
 3 & 39 & 89 & 0 & 0 & 0
 \\
 \hline
 1 & 39  & 88 & 1 & 0 & 1
 \\
 \hline
 1 & 39  & 88 & 1 & 0 & 0
 \\
 \hline
 2 & 39 & 88  & 0 & 1 & 0 
 \\ 
 \hline
 2 & 39 & 88 & 0 & 0 & 1
 \\ 
 \hline
 1 & 39 & 88 & 0 & 0 & 0
 \\ 
 \hline
 3 & 39 & 87 & 0 & 2 & 0
 \\ 
  \hline
 2 & 39 & 87 &  0 & 1 & 0
  \\
 \hline
 1  & 39 & 87 & 0 & 0 & 0
 \\ 
 \hline
 2 & 39 & 86 & 0 & 2 & 0 
 \\ 
 \hline
 2 & 39 & 86 & 1 & 0 & 0
 \\ 
 \hline
 4 & 39 & 86 & 0 & 1 & 0
 \\ 
 \hline
 7 & 39 & 86 & 0 & 0 & 0
 \\ 
 \hline
\end{tabular}
\end{center}
\caption{A summary of the different flux choices with $N_{\rm flux}=39$ that we have analyzed. The first column lists the number of $S_6$ distinct solutions that exhibit the shown pattern of massive fields and massless stabilized fields at the $r$-th power in the superpotential. It is of course possible that these different flux choices behave differently and split further if one were to go to even higher powers in the superpotential. \label{tab:N39}}
\end{table}

\subsection{Fully stabilized Minkowski vacua}
Lastly, we present two solutions that were part of our randomly generated set of solutions with $N_{\rm flux}=40$ and that actually do stabilize \emph{all} fields through higher order terms. Given that there are only 2 out of 142 examples, this shows that these are rare but they do exist. These two solutions provide to our knowledge the first Minkowski vacua without any flat directions (besides 11d supergravity). They both have 86 massive fields, 0 stabilized fields when including cubic terms via $W_3$, and finally 5 stabilized fields when including $W_4$. At that point, we do not need to worry about higher-order terms since we cannot stabilize any more fields and $\partial_i W_n=0$ is automatically satisfied for all $n$ since all moduli are stabilized at the origin of moduli space. Note that these solutions have $N_{\rm flux}=40$ and 91 stabilized fields. They are thus violating the Refined Tadpole Conjecture that would allow only the stabilization of less than $40\cdot\frac32=60$ fields for $N_{\rm flux}=40$.

The two solutions are 
\be \label{eq:BigOmega2}
\begin{split}
G^{15} &= \frac{1}{32} \left[ \left(-\frac{1}{4}+\frac{\rmi}{4}\right) \Omega_{1,1,2,3,1,2}+\left(\frac{1}{4}+\frac{\rmi}{4}\right) \Omega_{1,2,2,1,3,1}+\left(\frac{1}{4}-\frac{\rmi}{4}\right) \Omega_{1,2,3,2,1,1} \right.\\
&\qquad-\left(\frac{1}{4}+\frac{\rmi}{4}\right) \Omega_{1,3,1,2,1,2}+\frac{1}{2} \Omega_{1,3,2,1,1,2}-\frac{1}{2} \Omega_{1,3,2,1,2,1}\\
&\qquad +\left(\frac{1}{4}-\frac{\rmi}{4}\right) \Omega_{2,1,1,3,1,2}-\left(\frac{1}{4}+\frac{\rmi}{4}\right) \Omega_{2,1,2,1,3,1}+\frac{1}{4} \Omega_{2,1,2,2,1,2}\\
&\qquad +\left(\frac{1}{4}-\frac{\rmi}{4}\right) \Omega_{2,1,2,3,1,1}-\left(\frac{1}{4}-\frac{\rmi}{4}\right) \Omega_{2,1,3,2,1,1}-\left(\frac{1}{4}+\frac{\rmi}{4}\right) \Omega_{2,2,1,1,3,1}\\
&\qquad+\frac{\rmi}{4} \Omega_{2,2,1,2,1,2}-\left(\frac{1}{4}+\frac{\rmi}{4}\right) \Omega_{2,2,2,1,1,2}+\frac{1}{4} \Omega_{2,2,2,1,2,1}\\
&\qquad-\frac{1}{4} \Omega_{2,2,2,2,1,1}-\left(\frac{1}{4}-\frac{\rmi}{4}\right) \Omega_{2,2,3,1,1,1}-\left(\frac{1}{4}-\frac{\rmi}{4}\right) \Omega_{2,3,1,1,1,2}\\
&\qquad+\frac{1}{2} \Omega_{2,3,1,1,2,1}+\left(\frac{1}{4}+\frac{\rmi}{4}\right) \Omega_{2,3,1,2,1,1}+\left(\frac{1}{2}+\frac{\rmi}{2}\right) \Omega_{3,1,1,1,3,1}\\
&\qquad-\left(\frac{1}{4}+\frac{\rmi}{4}\right) \Omega_{3,1,1,2,1,2}-\left(\frac{1}{2}-\frac{\rmi}{2}\right) \Omega_{3,1,1,3,1,1}-\left(\frac{1}{4}-\frac{\rmi}{4}\right) \Omega_{3,1,2,1,1,2}\\
&\qquad+\left(\frac{1}{2}-\frac{\rmi}{2}\right) \Omega_{3,1,3,1,1,1}+\frac{1}{2} \Omega_{3,2,1,1,1,2}-\frac{1}{2} \Omega_{3,2,1,1,2,1}\\
&\qquad \left. +\left(\frac{1}{4}-\frac{\rmi}{4}\right) \Omega_{3,2,1,2,1,1}+\left(\frac{1}{4}+\frac{\rmi}{4}\right) \Omega_{3,2,2,1,1,1}-\left(\frac{1}{2}+\frac{\rmi}{2}\right) \Omega_{3,3,1,1,1,1} \right]
\end{split}
\ee
and
\begin{align} \label{eq:BigOmega3}
G^{16} &= \frac{1}{32} \left[ \left(\frac{1}{4}+\frac{\rmi}{4}\right) \Omega_{1,1,2,3,2,1}+\left(\frac{1}{4}-\frac{\rmi}{4}\right) \Omega_{1,2,2,1,1,3}-\left(\frac{1}{4}+\frac{\rmi}{4}\right) \Omega_{1,2,3,2,1,1} \right.\cr
&\qquad-\left(\frac{1}{4}-\frac{\rmi}{4}\right) \Omega_{1,3,1,2,2,1}+\frac{\rmi}{2} \Omega_{1,3,2,1,1,2}-\frac{\rmi}{2} \Omega_{1,3,2,1,2,1}\cr
&\qquad-\left(\frac{1}{4}+\frac{\rmi}{4}\right) \Omega_{2,1,1,3,2,1}-\left(\frac{1}{4}-\frac{\rmi}{4}\right) \Omega_{2,1,2,1,1,3}-\frac{\rmi}{4} \Omega_{2,1,2,2,2,1}\cr
&\qquad-\left(\frac{1}{4}+\frac{\rmi}{4}\right) \Omega_{2,1,2,3,1,1}+\left(\frac{1}{4}+\frac{\rmi}{4}\right) \Omega_{2,1,3,2,1,1}-\left(\frac{1}{4}-\frac{\rmi}{4}\right) \Omega_{2,2,1,1,1,3}\cr
&\qquad+\frac{1}{4} \Omega_{2,2,1,2,2,1}-\frac{\rmi}{4} \Omega_{2,2,2,1,1,2}-\left(\frac{1}{4}-\frac{\rmi}{4}\right) \Omega_{2,2,2,1,2,1}\\
&\qquad+\frac{\rmi}{4} \Omega_{2,2,2,2,1,1}+\left(\frac{1}{4}+\frac{\rmi}{4}\right) \Omega_{2,2,3,1,1,1}-\frac{\rmi}{2} \Omega_{2,3,1,1,1,2}\cr
&\qquad+\left(\frac{1}{4}+\frac{\rmi}{4}\right) \Omega_{2,3,1,1,2,1}+\left(\frac{1}{4}-\frac{\rmi}{4}\right) \Omega_{2,3,1,2,1,1}+\left(\frac{1}{2}-\frac{\rmi}{2}\right) \Omega_{3,1,1,1,1,3}\cr
&\qquad-\left(\frac{1}{4}-\frac{\rmi}{4}\right) \Omega_{3,1,1,2,2,1}+\left(\frac{1}{2}+\frac{\rmi}{2}\right) \Omega_{3,1,1,3,1,1}+\left(\frac{1}{4}+\frac{\rmi}{4}\right) \Omega_{3,1,2,1,2,1}\cr
&\qquad-\left(\frac{1}{2}+\frac{\rmi}{2}\right) \Omega_{3,1,3,1,1,1}+\frac{\rmi}{2} \Omega_{3,2,1,1,1,2}-\frac{\rmi}{2} \Omega_{3,2,1,1,2,1}\cr
&\qquad\left. -\left(\frac{1}{4}+\frac{\rmi}{4}\right) \Omega_{3,2,1,2,1,1}+\left(\frac{1}{4}-\frac{\rmi}{4}\right) \Omega_{3,2,2,1,1,1}-\left(\frac{1}{2}-\frac{\rmi}{2}\right) \Omega_{3,3,1,1,1,1} \right]\,.\nonumber
\end{align}

In a paper that appeared at the same time as ours \cite{Becker:2024ayh}, the authors found many flux configurations in this model that have $N_{\rm flux}=40$ and full rank 91 for the mass matrix. So, these solutions in \cite{Becker:2024ayh} are the first Minkowski vacua in which all scalar fields are massive. We comment further on the significance of these and our solutions in the next section.

\section{Implications for the swampland and the landscape}\label{sec:swamp}
In this section, we discuss the implications of the different flux compactifications of the $2^6$ LG model for our understanding of the landscape and the swampland conjectures.

\subsection{Full moduli stabilization and 4d $\mathcal{N}=1$ vacua}
Full moduli stabilization, i.e. finding \emph{isolated} vacua in the string landscape, is arguably one of the greatest challenges in string phenomenology today. Roughly twenty years ago many scenarios were put forward like KKLT \cite{Kachru:2003aw}, LVS \cite{Balasubramanian:2005zx, Conlon:2005ki} and DGKT \cite{DeWolfe:2005uu} to name a few. However, throughout the last decade, all of these constructions have been scrutinized and certain aspects have been called into question within the swampland program. This means that there are currently no constructions that are agreed upon by the entire string phenomenology community which leaves us in a dire spot. Note that this is not related to the presence or absence of supersymmetry or the value of the cosmological constant.

Finding new simple models with full moduli stabilization has thus become an important goal in string phenomenology. Non-geometric LG models provide an intriguing way of doing so. On the one hand, they allow us to bring the full strength of string theory to the problem. On the other hand, they give rise to very simple 4d $\mathcal{N}=1$ supergravity theories that depend, in the presence of fluxes, on all moduli. The initial study of the simplest of these models, the $1^9$ model, has discovered the presence of many massless fields and potential directions that do not get lifted even at higher order \cite{Becker:2024ijy}. However, the next more complicated example, the $2^6$ model that is studied here and at the same time in \cite{Becker:2024ayh} has completely changed the picture. Fully stabilized, isolated Minkowski vacua have been found in this paper albeit with some massless fields. The authors of \cite{Becker:2024ayh} found isolated Minkowski vacua without massless fields. Thus, there are now new constructions of moduli stabilization for the community to understand better and scrutinize in the future.

Let us already provide here a few comments on why we believe this moduli stabilization scenario is trustworthy. We have solved the equations $W=\partial_iW=0$, which do not depend on the K\"ahler potential at all but only on the superpotential $W$. For non-geometric LG models it was shown in \cite{Becker:2006ks} that the superpotential is exact and does not receive any perturbative or non-perturbative corrections. For example, the absence of the standard ED3 instanton corrections follows from $h^{1,1}=0$, i.e., the absence of 4-cycles. For D(-1) instantons it was likewise argued that they vanish in these models. This is consistent with the recent results in \cite{Kim:2022jvv}, where it was shown that they are only non-zero in models where the D7-brane charge is not canceled locally. Since the LG models have no D7-branes or O7-planes there is no such charge at all. Thus, the superpotential does not receive any corrections at all and any Minkowski vacuum is actually trustworthy. This reasoning applies to the two solutions above in equations \eqref{eq:BigOmega2} and \eqref{eq:BigOmega3} that have 86 massive and 5 massless but stabilized scalar fields. It also applies equally to the Minkowski vacua found in the paper \cite{Becker:2024ayh} that have 91 massive fields and therefore do not require any stabilization at higher order.

\subsection{The Massless Minkowski Conjecture}
\label{sec:massless-mink}
The Massless Minkowski Conjecture states that 10d supergravity compactifications to 4d Minkowski space always have a massless scalar field \cite{Andriot:2022yyj}. Here we are not dealing with compactifications of type IIB supergravity so we might be violating an important assumption of the conjecture. So, strictly speaking, the LG models cannot violate this conjecture. However, let us nevertheless entertain the possibility that all Minkowski vacua have a massless scalar field.

To our knowledge, the only known Minkowski vacuum without a massless scalar field was 11d supergravity or M-theory. When talking about compactifications to lower dimensions, one necessarily faces the challenges of giving masses to all the resulting scalar fields. As discussed in the previous subsection, moduli stabilization is hard and the most trustworthy (simple) models could therefore always contain massless fields. The presence of massless fields would then be a lamppost effect and by studying more complicated setups one would be able to eventually find fully stabilized models.\footnote{We are \emph{not} implying that the Massless Minkowski Conjecture is the result of a lamppost effect. It is motivated by a universal tachyon in certain dS solutions \cite{Danielsson:2012et}. This tachyon becomes massless when sending the cosmological constant to zero.}

One can then set out to find more and more intricate constructions that eventually lead to Minkowski vacua without flat directions. Before doing so one would however have to contemplate the following problem, which we spell out for 4d solutions. If one allows for $\mathcal{N}\geq 2$, then vector multiplets contain scalar fields as well. Large amounts of supersymmetry do protect the corresponding moduli space from perturbative and even non-perturbative corrections and we can fully trust these solutions but there is no clear way of having no massless scalar fields at all. One intriguing idea that was pursued recently in \cite{Baykara:2023plc} is the study of asymmetric orbifolds that give rise to models with no hypermultiplets. If among those one would find models without vector multiplets, i.e. theories that are pure supergravities then one would have succeeded. This has not yet been possible although the existence of a potential pure supergravity in 5d is discussed in that paper. With less or no supersymmetry, one faces the problem of perturbative and non-perturbative corrections. For example, even in 4d theories with $\mathcal{N}=1$, the superpotential is not protected from receiving corrections. Thus, there is usually an infinite number of unknown corrections that can affect any Minkowski vacuum. If these corrections would change the value of the scalar potential at the minimum away from zero, then the solution would be no longer a Minkowski vacuum. Stated differently, for $\mathcal{N}\leq 1$ an infinite number of corrections would have to sum to zero, making Minkowski vacua seemingly infinitely fine-tuned.

As we discussed in the previous subsection, there are no perturbative or non-perturbative corrections for these non-geometric LG models. Thus, due to these powerful non-renormalization theorems, we do not have to worry about infinite numbers of corrections and can trust the fully stabilized Minkowski vacua we found in this paper. This reasoning of course also applies to the Minkowski vacua found in the paper \cite{Becker:2024ayh} that have 91 massive fields and do not even require any stabilization at higher order.

\subsection{The Tadpole Conjecture}
The Tadpole Conjecture postulates a surprising relation between the length of an ISD flux $G\in H^{2,1}$ and the number $n_{\rm stab}$ of complex structure moduli that the flux lifts. In particular, it states that 
\be\label{eq:tadpoleconj}
N_{\rm flux} = \frac{1}{\tau-\bar{\tau}}\int G_3 \w \bar{G}_3 > 2\,\alpha \, n_{\rm stab}\,,
\ee
with $\alpha >\tfrac13$ in its original refined formulation \cite{Bena:2020xrh}. While this was not explicitly specified it probably makes sense to take $n_{\rm stab}$ here to be the total number of stabilized fields and not just the number of massive fields $n_{\rm massive}$. However, in principle, one could also entertain the conjecture
\be
N_{\rm flux} = \frac{1}{\tau-\bar{\tau}}\int G_3 \w \bar{G}_3 >2\, \alpha \, n_{\rm massive}\,,
\ee
which is weaker since $n_{\rm stab} \geq n_{\rm massive}$. 

Either form of the conjecture is violated by our flux choice above in equation \eqref{eq:TwoOmega1} by more than a factor of 2, which seemed to be a previously found weaker bound in the $1^9$ LG model \cite{Becker:2024ijy}. That flux choice in equation \eqref{eq:TwoOmega1} had $N_{\rm flux}=16$ and 52 massive fields, which gives
\be
\frac{N_{\rm flux}}{n_{\rm massive}} \approx .308 < \frac13\,.
\ee
One might thus be tempted to conclude that the Tadpole Conjecture is not correct or does not apply to this $2^6$ LG model. 

However, the authors of \cite{Becker:2024ijy} generated a large set of flux configurations in the $1^9$ LG model and found a confirmation of the Tadpole Conjecture albeit with a different factor (see figure 2 on page 22 of that paper). A similar study for the $2^6$ LG model was undertaken in the paper \cite{Becker:2024ayh} that appeared at the same time as ours. The authors kindly shared their results with us and those do show a confirmation of a bound like the one in equation \eqref{eq:tadpoleconj} albeit with a much smaller coefficient. Interestingly, the authors of \cite{Becker:2024ayh} find $\alpha \lesssim 1/6.8$ which is smaller than the F-theory expectation $\alpha=1/4$.

\section{Conclusion}\label{sec:conclusion} 
In this paper, we have studied flux compactifications on an orientifold of the $2^6$ Landau-Ginzburg model. We focused on 4d $\mathcal{N}=1$ Minkowski vacua that arise for $G_3$ fluxes that are imaginary-self-dual. We demonstrated that, for several flux choices, there is non-trivial moduli stabilization from higher-than-quadratic order terms in the expansion of the superpotential around a critical point. This means that there are massless fields that are stabilized by higher order interaction terms, similar to a massless $\phi^4$ scalar field that has no flat direction. Unlike in the $1^9$ Landau-Ginzburg model \cite{Becker:2024ijy}, it is not tractable to study all flux examples beyond terms in the superpotential that are cubic in the complex structure moduli. However, we find two interesting solutions in the $2^6$ model that violate the Refined Tadpole Conjecture. One example has 52 massive fields for a flux choice that only contributes $N_{\rm flux}=16$ to the O3 tadpole condition. Another example has $N_{\rm flux}=28$, and 75 massive fields but has at least $84=3\cdot28$ stabilized fields. Both examples violates the Refined Tadpole Conjecture by more than a factor of 2. Nevertheless, it is possible that this model is actually confirming the original Tadpole Conjecture, with a modified bound, as is shown by studying a large number of flux configurations in \cite{Becker:2024ayh}.

Even more interestingly, we provide examples where supersymmetric Minkowski critical points without full mass matrix rank turn out to be isolated vacua once we include higher order terms in the superpotential. Specifically, we have two flux choices for which 86 out of the 91 fields are massive and the 5 remaining fields get stabilized at higher order. Due to the non-renormalization theorem for the superpotential $W$ that was derived in \cite{Becker:2006ks}, these isolated supersymmetric Minkowski vacua should be fully controlled and are arguably the simplest four-dimensional solutions of string theory without a moduli space.

\section*{Acknowledgments}
We like to thank Katrin Becker, Mariana Gra\~na, Daniel Junghans, Miguel Montero, Miguel Morros, Sav Sethi, Johannes Walcher, and Qi You for useful discussions.
The work of AS is supported in part by the NSF grant PHY-2112859. MR acknowledges the support of the Dr. Hyo Sang Lee Graduate Fellowship from the College of Arts and Sciences at Lehigh University. The work of MR and TW is supported in part by the NSF grant PHY-2210271. MR and TW would like to thank the Erwin-Schr\"odinger-Institute in Vienna for their hospitality; this work was completed there during the program ``The Landscape vs. the Swampland".
\appendix

\newpage

\bibliographystyle{JHEP}
\bibliography{refs}

\providecommand{\href}[2]{#2}\begingroup\raggedright\begin{thebibliography}{10}

\bibitem{DeWolfe:2005uu}
O.~DeWolfe, A.~Giryavets, S.~Kachru and W.~Taylor, \emph{{Type IIA moduli
  stabilization}},
  \href{https://doi.org/10.1088/1126-6708/2005/07/066}{\emph{JHEP} {\bfseries
  07} (2005) 066} [\href{https://arxiv.org/abs/hep-th/0505160}{{\ttfamily
  hep-th/0505160}}].

\bibitem{Giddings:2001yu}
S.B.~Giddings, S.~Kachru and J.~Polchinski, \emph{{Hierarchies from fluxes in
  string compactifications}},
  \href{https://doi.org/10.1103/PhysRevD.66.106006}{\emph{Phys. Rev. D}
  {\bfseries 66} (2002) 106006}
  [\href{https://arxiv.org/abs/hep-th/0105097}{{\ttfamily hep-th/0105097}}].

\bibitem{Junghans:2020acz}
D.~Junghans, \emph{{O-Plane Backreaction and Scale Separation in Type IIA Flux
  Vacua}}, \href{https://doi.org/10.1002/prop.202000040}{\emph{Fortsch. Phys.}
  {\bfseries 68} (2020) 2000040}
  [\href{https://arxiv.org/abs/2003.06274}{{\ttfamily 2003.06274}}].

\bibitem{Marchesano:2020qvg}
F.~Marchesano, E.~Palti, J.~Quirant and A.~Tomasiello, \emph{{On supersymmetric
  AdS$_{4}$ orientifold vacua}},
  \href{https://doi.org/10.1007/JHEP08(2020)087}{\emph{JHEP} {\bfseries 08}
  (2020) 087} [\href{https://arxiv.org/abs/2003.13578}{{\ttfamily
  2003.13578}}].

\bibitem{Cribiori:2021djm}
N.~Cribiori, D.~Junghans, V.~Van~Hemelryck, T.~Van~Riet and T.~Wrase,
  \emph{{Scale-separated AdS4 vacua of IIA orientifolds and M-theory}},
  \href{https://doi.org/10.1103/PhysRevD.104.126014}{\emph{Phys. Rev. D}
  {\bfseries 104} (2021) 126014}
  [\href{https://arxiv.org/abs/2107.00019}{{\ttfamily 2107.00019}}].

\bibitem{Emelin:2022cac}
M.~Emelin, F.~Farakos and G.~Tringas, \emph{{O6-plane backreaction on
  scale-separated Type IIA AdS$_{3}$ vacua}},
  \href{https://doi.org/10.1007/JHEP07(2022)133}{\emph{JHEP} {\bfseries 07}
  (2022) 133} [\href{https://arxiv.org/abs/2202.13431}{{\ttfamily
  2202.13431}}].

\bibitem{Junghans:2023yue}
D.~Junghans, \emph{{A note on O6 intersections in AdS flux vacua}},
  \href{https://doi.org/10.1007/JHEP02(2024)126}{\emph{JHEP} {\bfseries 02}
  (2024) 126} [\href{https://arxiv.org/abs/2310.17695}{{\ttfamily
  2310.17695}}].

\bibitem{Bardzell:2024anh}
J.~Bardzell, K.~Federico, D.~Smith and T.~Wrase, \emph{{On the absence of
  supergravity solutions for localized, intersecting sources}},
  \href{https://doi.org/10.1007/JHEP06(2024)083}{\emph{JHEP} {\bfseries 06}
  (2024) 083} [\href{https://arxiv.org/abs/2403.09873}{{\ttfamily
  2403.09873}}].

\bibitem{Emelin:2024vug}
M.~Emelin, \emph{{Consistency conditions for O-plane unsmearing from
  second-order perturbation theory}},
  \href{https://arxiv.org/abs/2407.12717}{{\ttfamily 2407.12717}}.

\bibitem{Kachru:2003aw}
S.~Kachru, R.~Kallosh, A.D.~Linde and S.P.~Trivedi, \emph{{De Sitter vacua in
  string theory}},
  \href{https://doi.org/10.1103/PhysRevD.68.046005}{\emph{Phys. Rev. D}
  {\bfseries 68} (2003) 046005}
  [\href{https://arxiv.org/abs/hep-th/0301240}{{\ttfamily hep-th/0301240}}].

\bibitem{Becker:2006ks}
K.~Becker, M.~Becker, C.~Vafa and J.~Walcher, \emph{{Moduli Stabilization in
  Non-Geometric Backgrounds}},
  \href{https://doi.org/10.1016/j.nuclphysb.2007.01.034}{\emph{Nucl. Phys. B}
  {\bfseries 770} (2007) 1}
  [\href{https://arxiv.org/abs/hep-th/0611001}{{\ttfamily hep-th/0611001}}].

\bibitem{Vafa:1988uu}
C.~Vafa and N.P.~Warner, \emph{{Catastrophes and the Classification of
  Conformal Theories}},
  \href{https://doi.org/10.1016/0370-2693(89)90473-5}{\emph{Phys. Lett. B}
  {\bfseries 218} (1989) 51}.

\bibitem{vafaOrbifoldized}
C.~Vafa, \emph{{String Vacua and Orbifoldized L-G Models}},
  \href{https://doi.org/10.1142/S0217732389001350}{\emph{Mod. Phys. Lett. A}
  {\bfseries 4} (1989) 1169}.

\bibitem{Schimmrigk:1992ai}
R.~Schimmrigk, \emph{{Critical superstring vacua from noncritical manifolds: A
  Novel framework for string compactification}},
  \href{https://doi.org/10.1103/PhysRevLett.70.3688}{\emph{Phys. Rev. Lett.}
  {\bfseries 70} (1993) 3688}
  [\href{https://arxiv.org/abs/hep-th/9210062}{{\ttfamily hep-th/9210062}}].

\bibitem{Witten:1993yc}
E.~Witten, \emph{{Phases of N=2 theories in two-dimensions}},
  \href{https://doi.org/10.1016/0550-3213(93)90033-L}{\emph{Nucl. Phys. B}
  {\bfseries 403} (1993) 159}
  [\href{https://arxiv.org/abs/hep-th/9301042}{{\ttfamily hep-th/9301042}}].

\bibitem{Candelas:1993nd}
P.~Candelas, E.~Derrick and L.~Parkes, \emph{{Generalized Calabi-Yau manifolds
  and the mirror of a rigid manifold}},
  \href{https://doi.org/10.1016/0550-3213(93)90276-U}{\emph{Nucl. Phys. B}
  {\bfseries 407} (1993) 115}
  [\href{https://arxiv.org/abs/hep-th/9304045}{{\ttfamily hep-th/9304045}}].

\bibitem{Becker:2007dn}
K.~Becker, M.~Becker and J.~Walcher, \emph{{Runaway in the Landscape}},
  \href{https://doi.org/10.1103/PhysRevD.76.106002}{\emph{Phys. Rev. D}
  {\bfseries 76} (2007) 106002}
  [\href{https://arxiv.org/abs/0706.0514}{{\ttfamily 0706.0514}}].

\bibitem{Becker:2007ee}
K.~Becker, Y.-C.~Chung and G.-Y.~Guo, \emph{{Metastable Flux Configurations and
  de Sitter Spaces}},
  \href{https://doi.org/10.1016/j.nuclphysb.2007.09.019}{\emph{Nucl. Phys. B}
  {\bfseries 790} (2008) 240}
  [\href{https://arxiv.org/abs/0706.2502}{{\ttfamily 0706.2502}}].

\bibitem{Ishiguro:2021csu}
K.~Ishiguro and H.~Otsuka, \emph{{Sharpening the boundaries between flux
  landscape and swampland by tadpole charge}},
  \href{https://doi.org/10.1007/JHEP12(2021)017}{\emph{JHEP} {\bfseries 12}
  (2021) 017} [\href{https://arxiv.org/abs/2104.15030}{{\ttfamily
  2104.15030}}].

\bibitem{Bardzell:2022jfh}
J.~Bardzell, E.~Gonzalo, M.~Rajaguru, D.~Smith and T.~Wrase, \emph{{Type IIB
  flux compactifications with h$^{1,1}$ = 0}},
  \href{https://doi.org/10.1007/JHEP06(2022)166}{\emph{JHEP} {\bfseries 06}
  (2022) 166} [\href{https://arxiv.org/abs/2203.15818}{{\ttfamily
  2203.15818}}].

\bibitem{Becker:2022hse}
K.~Becker, E.~Gonzalo, J.~Walcher and T.~Wrase, \emph{{Fluxes, vacua, and
  tadpoles meet Landau-Ginzburg and Fermat}},
  \href{https://doi.org/10.1007/JHEP12(2022)083}{\emph{JHEP} {\bfseries 12}
  (2022) 083} [\href{https://arxiv.org/abs/2210.03706}{{\ttfamily
  2210.03706}}].

\bibitem{Cremonini:2023suw}
S.~Cremonini, E.~Gonzalo, M.~Rajaguru, Y.~Tang and T.~Wrase, \emph{{On
  asymptotic dark energy in string theory}},
  \href{https://doi.org/10.1007/JHEP09(2023)075}{\emph{JHEP} {\bfseries 09}
  (2023) 075} [\href{https://arxiv.org/abs/2306.15714}{{\ttfamily
  2306.15714}}].

\bibitem{Becker:2023rqi}
K.~Becker, N.~Brady and A.~Sengupta, \emph{{On fluxes in the 1$^{9}$
  Landau-Ginzburg model}},
  \href{https://doi.org/10.1007/JHEP11(2023)152}{\emph{JHEP} {\bfseries 11}
  (2023) 152} [\href{https://arxiv.org/abs/2310.00770}{{\ttfamily
  2310.00770}}].

\bibitem{Becker:2024ijy}
K.~Becker, M.~Rajaguru, A.~Sengupta, J.~Walcher and T.~Wrase,
  \emph{{Stabilizing massless fields with fluxes in Landau-Ginzburg models}},
  \href{https://arxiv.org/abs/2406.03435}{{\ttfamily 2406.03435}}.

\bibitem{Ishiguro:2024coq}
K.~Ishiguro, T.~Kai and H.~Otsuka, \emph{{Stabilization of a twisted modulus on
  a mirror of rigid Calabi-Yau manifold}},
  \href{https://arxiv.org/abs/2406.08970}{{\ttfamily 2406.08970}}.

\bibitem{Giryavets:2003vd}
A.~Giryavets, S.~Kachru, P.K.~Tripathy and S.P.~Trivedi, \emph{{Flux
  compactifications on Calabi-Yau threefolds}},
  \href{https://doi.org/10.1088/1126-6708/2004/04/003}{\emph{JHEP} {\bfseries
  04} (2004) 003} [\href{https://arxiv.org/abs/hep-th/0312104}{{\ttfamily
  hep-th/0312104}}].

\bibitem{Denef:2004dm}
F.~Denef, M.R.~Douglas and B.~Florea, \emph{{Building a better racetrack}},
  \href{https://doi.org/10.1088/1126-6708/2004/06/034}{\emph{JHEP} {\bfseries
  06} (2004) 034} [\href{https://arxiv.org/abs/hep-th/0404257}{{\ttfamily
  hep-th/0404257}}].

\bibitem{Denef:2005mm}
F.~Denef, M.R.~Douglas, B.~Florea, A.~Grassi and S.~Kachru, \emph{{Fixing all
  moduli in a simple f-theory compactification}},
  \href{https://doi.org/10.4310/ATMP.2005.v9.n6.a1}{\emph{Adv. Theor. Math.
  Phys.} {\bfseries 9} (2005) 861}
  [\href{https://arxiv.org/abs/hep-th/0503124}{{\ttfamily hep-th/0503124}}].

\bibitem{Braun:2020jrx}
A.P.~Braun and R.~Valandro, \emph{{$G_{4}$ flux, algebraic cycles and complex
  structure moduli stabilization}},
  \href{https://doi.org/10.1007/JHEP01(2021)207}{\emph{JHEP} {\bfseries 01}
  (2021) 207} [\href{https://arxiv.org/abs/2009.11873}{{\ttfamily
  2009.11873}}].

\bibitem{Bena:2020xrh}
I.~Bena, J.~Bl\r{a}b\"ack, M.~Gra\~na and S.~L\"ust, \emph{{The tadpole
  problem}}, \href{https://doi.org/10.1007/JHEP11(2021)223}{\emph{JHEP}
  {\bfseries 11} (2021) 223}
  [\href{https://arxiv.org/abs/2010.10519}{{\ttfamily 2010.10519}}].

\bibitem{Bena:2021wyr}
I.~Bena, J.~Bl\r{a}b\"ack, M.~Gra\~na and S.~L\"ust, \emph{{Algorithmically
  Solving the Tadpole Problem}},
  \href{https://doi.org/10.1007/s00006-021-01189-6}{\emph{Adv. Appl. Clifford
  Algebras} {\bfseries 32} (2022) 7}
  [\href{https://arxiv.org/abs/2103.03250}{{\ttfamily 2103.03250}}].

\bibitem{Marchesano:2021gyv}
F.~Marchesano, D.~Prieto and M.~Wiesner, \emph{{F-theory flux vacua at large
  complex structure}},
  \href{https://doi.org/10.1007/JHEP08(2021)077}{\emph{JHEP} {\bfseries 08}
  (2021) 077} [\href{https://arxiv.org/abs/2105.09326}{{\ttfamily
  2105.09326}}].

\bibitem{Plauschinn:2021hkp}
E.~Plauschinn, \emph{{The tadpole conjecture at large complex-structure}},
  \href{https://doi.org/10.1007/JHEP02(2022)206}{\emph{JHEP} {\bfseries 02}
  (2022) 206} [\href{https://arxiv.org/abs/2109.00029}{{\ttfamily
  2109.00029}}].

\bibitem{Lust:2021xds}
S.~L\"ust, \emph{{Large complex structure flux vacua of IIB and the Tadpole
  Conjecture}},  \href{https://arxiv.org/abs/2109.05033}{{\ttfamily
  2109.05033}}.

\bibitem{Bena:2021qty}
I.~Bena, C.~Brodie and M.~Gra\~na, \emph{{D7 moduli stabilization: the tadpole
  menace}}, \href{https://doi.org/10.1007/JHEP01(2022)138}{\emph{JHEP}
  {\bfseries 01} (2022) 138}
  [\href{https://arxiv.org/abs/2112.00013}{{\ttfamily 2112.00013}}].

\bibitem{Grana:2022dfw}
M.~Gra\~na, T.W.~Grimm, D.~van~de Heisteeg, A.~Herraez and E.~Plauschinn,
  \emph{{The tadpole conjecture in asymptotic limits}},
  \href{https://doi.org/10.1007/JHEP08(2022)237}{\emph{JHEP} {\bfseries 08}
  (2022) 237} [\href{https://arxiv.org/abs/2204.05331}{{\ttfamily
  2204.05331}}].

\bibitem{Tsagkaris:2022apo}
K.~Tsagkaris and E.~Plauschinn, \emph{{Moduli stabilization in type IIB
  orientifolds at h$^{2,1}$ = 50}},
  \href{https://doi.org/10.1007/JHEP03(2023)049}{\emph{JHEP} {\bfseries 03}
  (2023) 049} [\href{https://arxiv.org/abs/2207.13721}{{\ttfamily
  2207.13721}}].

\bibitem{Lust:2022mhk}
S.~L\"ust and M.~Wiesner, \emph{{The tadpole conjecture in the interior of
  moduli space}}, \href{https://doi.org/10.1007/JHEP12(2023)029}{\emph{JHEP}
  {\bfseries 12} (2023) 029}
  [\href{https://arxiv.org/abs/2211.05128}{{\ttfamily 2211.05128}}].

\bibitem{Coudarchet:2023mmm}
T.~Coudarchet, F.~Marchesano, D.~Prieto and M.A.~Urkiola, \emph{{Symmetric
  fluxes and small tadpoles}},
  \href{https://doi.org/10.1007/JHEP08(2023)016}{\emph{JHEP} {\bfseries 08}
  (2023) 016} [\href{https://arxiv.org/abs/2304.04789}{{\ttfamily
  2304.04789}}].

\bibitem{Braun:2023pzd}
A.P.~Braun, B.~Fraiman, M.~Gra\~na, S.~L\"ust and H.~Parra~de Freitas,
  \emph{{Tadpoles and gauge symmetries}},
  \href{https://doi.org/10.1007/JHEP08(2023)134}{\emph{JHEP} {\bfseries 08}
  (2023) 134} [\href{https://arxiv.org/abs/2304.06751}{{\ttfamily
  2304.06751}}].

\bibitem{Braun:2023edp}
A.P.~Braun, H.~Fortin, D.L.~Garcia and R.V.~Loyola, \emph{{More on G-flux and
  general hodge cycles on the Fermat sextic}},
  \href{https://doi.org/10.1007/JHEP06(2024)046}{\emph{JHEP} {\bfseries 06}
  (2024) 046} [\href{https://arxiv.org/abs/2401.00470}{{\ttfamily
  2401.00470}}].

\bibitem{Becker:2024ayh}
K.~Becker, N.~Brady, M.~Gra\~na, M.~Morros, A.~Sengupta and Q.~You,
  \emph{{Tadpole conjecture in non-geometric backgrounds}},
  \href{https://arxiv.org/abs/2407.16758}{{\ttfamily 2407.16758}}.

\bibitem{Andriot:2022yyj}
D.~Andriot, L.~Horer and P.~Marconnet, \emph{{Exploring the landscape of
  (anti-) de Sitter and Minkowski solutions: group manifolds, stability and
  scale separation}},
  \href{https://doi.org/10.1007/JHEP08(2022)109}{\emph{JHEP} {\bfseries 08}
  (2022) 109} [\href{https://arxiv.org/abs/2204.05327}{{\ttfamily
  2204.05327}}].

\bibitem{Brunner:2004zd}
I.~Brunner, K.~Hori, K.~Hosomichi and J.~Walcher, \emph{{Orientifolds of Gepner
  models}}, \href{https://doi.org/10.1088/1126-6708/2007/02/001}{\emph{JHEP}
  {\bfseries 02} (2007) 001}
  [\href{https://arxiv.org/abs/hep-th/0401137}{{\ttfamily hep-th/0401137}}].

\bibitem{HoriIqbalVafa}
K.~Hori, A.~Iqbal and C.~Vafa, \emph{{D-branes and mirror symmetry}},
  \href{https://arxiv.org/abs/hep-th/0005247}{{\ttfamily hep-th/0005247}}.

\bibitem{Gukov:1999ya}
S.~Gukov, C.~Vafa and E.~Witten, \emph{{CFT's from Calabi-Yau four folds}},
  \href{https://doi.org/10.1016/S0550-3213(00)00373-4}{\emph{Nucl. Phys. B}
  {\bfseries 584} (2000) 69}
  [\href{https://arxiv.org/abs/hep-th/9906070}{{\ttfamily hep-th/9906070}}].

\bibitem{Grimm:2024fip}
T.W.~Grimm and D.~van~de Heisteeg, \emph{{Exact Flux Vacua, Symmetries, and the
  Structure of the Landscape}},
  \href{https://arxiv.org/abs/2404.12422}{{\ttfamily 2404.12422}}.

\bibitem{Balasubramanian:2005zx}
V.~Balasubramanian, P.~Berglund, J.P.~Conlon and F.~Quevedo, \emph{{Systematics
  of moduli stabilisation in Calabi-Yau flux compactifications}},
  \href{https://doi.org/10.1088/1126-6708/2005/03/007}{\emph{JHEP} {\bfseries
  03} (2005) 007} [\href{https://arxiv.org/abs/hep-th/0502058}{{\ttfamily
  hep-th/0502058}}].

\bibitem{Conlon:2005ki}
J.P.~Conlon, F.~Quevedo and K.~Suruliz, \emph{{Large-volume flux
  compactifications: Moduli spectrum and D3/D7 soft supersymmetry breaking}},
  \href{https://doi.org/10.1088/1126-6708/2005/08/007}{\emph{JHEP} {\bfseries
  08} (2005) 007} [\href{https://arxiv.org/abs/hep-th/0505076}{{\ttfamily
  hep-th/0505076}}].

\bibitem{Kim:2022jvv}
M.~Kim, \emph{{D-instanton superpotential in string theory}},
  \href{https://doi.org/10.1007/JHEP03(2022)054}{\emph{JHEP} {\bfseries 03}
  (2022) 054} [\href{https://arxiv.org/abs/2201.04634}{{\ttfamily
  2201.04634}}].

\bibitem{Danielsson:2012et}
U.H.~Danielsson, G.~Shiu, T.~Van~Riet and T.~Wrase, \emph{{A note on obstinate
  tachyons in classical dS solutions}},
  \href{https://doi.org/10.1007/JHEP03(2013)138}{\emph{JHEP} {\bfseries 03}
  (2013) 138} [\href{https://arxiv.org/abs/1212.5178}{{\ttfamily 1212.5178}}].

\bibitem{Baykara:2023plc}
Z.K.~Baykara, Y.~Hamada, H.-C.~Tarazi and C.~Vafa, \emph{{On the string
  landscape without hypermultiplets}},
  \href{https://doi.org/10.1007/JHEP04(2024)121}{\emph{JHEP} {\bfseries 04}
  (2024) 121} [\href{https://arxiv.org/abs/2309.15152}{{\ttfamily
  2309.15152}}].

\end{thebibliography}\endgroup

\end{document}